\documentclass[12pt,preprint]{aastex}
\slugcomment{Accepted and scheduled for publication  
in {\it The Astrophysical Journal},  for  the July 10, 2014, v 789, 2nd issue}  
\def\lax {\ifmmode{_<\atop^{\sim}}\else{${_<\atop^{\sim}}$}\fi}  
\def\gax {\ifmmode{_>\atop^{\sim}}\else{${_>\atop^{\sim}}$}\fi}  
\def\gtorder{\mathrel{\raise.3ex\hbox{$>$}\mkern-14mu
             \lower0.6ex\hbox{$\sim$}}}

\def\cm2{cm$^{-2}$}
\def\s1{s$^{-1}$}

\begin{document}


\title{X-ray spectral and timing behavior of Scorpius X-1. Spectral hardening   
during the flaring branch
}







\author{Lev Titarchuk\altaffilmark{1},  Elena Seifina\altaffilmark{2} \& Chris Shrader\altaffilmark{3, 4}}
\altaffiltext{1}{Dipartimento di Fisica, Universit\`a di Ferrara, Via Saragat 1, I-44122 Ferrara, Italy, email:titarchuk@fe.infn.it; 
George Mason University Fairfax, VA 22030;   
Goddard Space Flight Center, NASA,  code 663, Greenbelt  
MD 20770, USA; email:lev@milkyway.gsfc.nasa.gov, USA}
\altaffiltext{2}{Moscow M.V.~Lomonosov State University/Sternberg Astronomical Institute, Universitetsky 
Prospect 13, Moscow, 119992, Russia; seif@sai.msu.ru}
\altaffiltext{3}{NASA Goddard Space Flight Center, NASA, Astrophysics Science Division, Code 661, Greenbelt, MD 20771, USA; Chris.R.Shrader@nasa.gov}
\altaffiltext{4}{Universities Space Research Association, 10211 Wincopin Cir, Suite 500, Columbia, MD 21044, USA}

\begin{abstract}
We present an analysis of the spectral and timing properties 
of  X-ray emission from 
the {\it Z-}source 
Sco~X-1 
during its evolution  
between the  {\it Horizontal} (HB)  and {\it Flaring} (FB) branches 
observed with the {\it Rossi} X-ray Timing Explorer   during the 1996 -- 2002 period. 
We find that the  broad-band (3 -- 250 keV) 
energy spectra  during all 
spectral  states 
can be adequately reproduced by  a 
model, consisting   
of  
two Comptonized components  
and an iron-line. 
We suggest that 
the seed photons   of $kT_{s1}\lax$0.7 keV coming from the disk 
and 
of temperature $kT_{s2}\lax$1.8 keV coming from the neutron star (NS) 
are each  upscattered by 
hot electrons of a ``Compton cloud" (herein {\it Comptb1} and 
 {\it Comptb2} components respectively with which are associated  similarly subscripted parameters).   
The photon power-law index $\Gamma_{2}$ 
is almost constant ($\Gamma_{2}\sim 2$) for all spectral states.
 In  turn, 
$\Gamma_{1}$ 
demonstrates a two-phase behavior with the spectral state: 
$\Gamma_{1}$ is quasi-constant at the level $\Gamma_{1}\sim 2$ for the  HB$-$NB
and 
$\Gamma_{1}$ is  less than 2, namely in the range of  
$1.3<\Gamma_1<2$, when source traces 
the FB.
We also detect 
a decrease 
$kT_{s2}$ from 1.8 keV to 0.7 keV during the FB.
We 
interpret this apparent quasi-stability of the indices 
during the HB$-$NB  
in the framework of the model in which the spectrum 
is determined by 
the  Comptonized thermal 
components.
%
This 
established for the Comptonized spectral components  
of the {\it Z}-source Sco~X-1 
is similar to that was previously found in  the  {\it atoll} sources 4U~1728-34, GX~3+1 and 4U~1820-30 and  the {\it Z}-source 
GX~340+0 through all spectral states. 
However we interpret the index reduction phase detected during the FB in Sco~X-1 
 within the framework of a model in which the spectrum at the FB is determined by 
  high radiation pressure from the NS surface.

\end{abstract}


\section{Introduction}

The neutron stars (NSs) in X-ray binaries offer a unique opportunity to study  the properties 
of matter under the most extreme conditions.
These sources show a variety of observational manifestations which can be used to verify different  theoretical models. The {\it Z} sources being related 
to NSs  in X-ray binaries  radiating close to the Eddington limit ($L_{Edd}$), sometimes show   properties similar to black holes   (BHs).  
For example,  the {\it Z} source Scorpius~X-1,  hereafter Sco~X-1, 
exhibits  a strong  {\it hard tail} in its X-ray spectrum 
which for a long time  was considered to be a unique  BH signature \citep{roth80}. 

Sco~X-l and  other {\it Z} sources GX~349+2, GX~340+0,  
GX~17+2, GX~5-1 and Cyg~X-2 form 
a group with a similar behavior for  three spectral branches
(Hasinger \& van der Klis 1989). However there are  differences in the detailed observational behavior of these {\it Z} 
sources. As Sco~X-l and GX~17+2 and GX 349+2 show strong flares in the intensity  during the {\it flaring}  branches (FB), 
while GX~340+0, Cyg~X-2 and GX~5-1 undergo dips in the FB
(Hasinger \& van der Klis 1989, Penninx et al. 1991, Kuulkers et al. 1994). 
It is worth noting that the {\it normal} branch (NB) properties are similar for these two groups 
of {\it Z} sources, but {\it horizontal} branch (HB) and FB properties are fundamentally    different.

These 
distinct observational appearances  between 
these two groups are 
{
not fully understood.
}
{ 
An interesting suggestion 
has been made by Lin et al. (2009), hereafter LRH09, and Homan et al. (2010)  based on observations of the transient NS source XTE~J1701-462.  
They found that this source  showed characteristic features of a {\it Z} source,  and then  an {\it atoll} source when the luminosity 
 decreased. They claimed that the initial  Cyg-like behavior was
followed by Sco-like one when  the luminosity 
decreased  and then they  concluded  
 Cyg and Sco  types depended on  the luminosity. 
}

Sco X-1
is the brightest  persistent  X-ray source in the sky
and the first identified  as a X-ray extrasolar source (Giacconi et al.
1962). It is located at  a distance of 2.8$\pm$0.3 kpc (Bradshaw et al. 1999) and  its binary orbit 
has a low inclination 
angle to the Earth observer and thus is suitable for a study   of the {\it flaring} events in {\it Z}-sources. 
This binary system consists of an old,  weakly  magnetized neutron star with a
mass $\sim$1.4 $M_{\odot}$ in which accreting matter is transferred through Roche lobe overflow
from a low-mass companion (recently identified as a M-class
star of $\sim$0.4 M$_{\odot}$, see details in Steeghs \& Casares 2002).  Sco~X-1 is a 
prototype of the class of  low-mass X-ray binaries (LMXBs), 
which emit close to the Eddington limit for a 1.4 M$_{\odot}$
NS ($L_{Edd}=2\times 10^{38}$ ergs~s$^{-1}$).

Sco~X-1,   similar to 
other  {\it Z} sources, shows   quasi-periodic oscillations (QPOs) 
along all the branches of its {\it color-color diagram} (CCD). 
Extensive timing analysis of Sco~X-1 with EXOSAT indicate 
that the QPO phenomenon is closely related to the transition of Sco~X-1 between different spectral states
[e.g., van der Klis et al., 1987a,b; Middleditch \& Priedhorsky, 1986; 
Priedhorsky et al. 1986; Hasinger et al. 1989, hereafter 
HPM89, and 
Hertz et al., 1992). 
In particular, the transition between 
the HB $-$ NB and 
the FB states is characterized by variations in 
QPO frequency between 6 Hz and 16 Hz (with no intensity correlation). When Sco~X-1 enters into the prolongated 
NB state the QPO frequency settles  at 6 -- 8 Hz range (anti-correlated with intensity). Low frequency noise (LFN) component of  the power spectrum
becomes stronger when the source intensity  increases  at the FB and at the same time the QPOs disappear. 
In turn, Dieters \& van der Klis (2000)  using  EXOSAT, deteregard cted,  an abrupt increase of the QPO frequency from 6 Hz to 10 Hz, a so called 
{\it rapid~excursions} near {\it soft} apex in hardness-intensity 
diagram (HID) 
along with {\it grand} transitions from 8 Hz to 21 Hz  when the object went  from the {\it lower} FB to the {\it upper} FB. 

Based on  the {\it RXTE} observations van der Klis et al. (1996) reported  the discovery of 45 Hz QPOs, mostly prominent in the middle of the NB. 
Casella et al. (2006) detected a monotonic (smooth) increase 
of the QPO centroid frequency from 4.5 -- 7 Hz (at the NB) to 6 -- 25 Hz (at the FB), which can indicate  to 
the same nature of these low frequency QPOs. 
The power spectrum of Sco~X-1 also includes   a pair of QPOs, which frequencies are in the range 800 -- 1100 Hz, the so called kHz QPOs. 
The peak frequency separation   is weakly frequency-dependent  (Zhang et al. 2006).

The physical interpretation of these timing features is not unique, and the related scientific
debate is still open. 
Apart from detailed knowledge of 
timing properties of Sco~X-1, the spectral studies  have been 
not so extensive.
This is mostly caused 
by   the  brightness of Sco~X-1 that prevents this source from  
direct observation with the high resolution X-ray satellites, e.g., 
{\it Chandra} and XMM-{\it Newton}. However,
a significant 
progress was achieved in this regard 
due to the detailed 
investigations of broad-band spectra 
of Sco~X-1 based on {\it RXTE} observations
by 
 \cite{D'Amico01}, \cite{Barnard03}, 
\cite{Bradshaw03},  \cite{D'Ai07} and \cite{Church12}. These authors demonstrated 
that 
X-ray spectrum of Sco~X-1 is well fit
by a two-component model consisting of a {\it thermal}  component, related to the NS, and a { \it Comptonized} 
component, associated with an extended accretion disk corona. In addition, a broad emission line with a {\it Gaussian} profile is applied to this model in all spectral states.
However, the results of this analysis 
and their interpretation 
depend on the energy band considered and the 
adopted spectral model 
for the {\it  thermal }  and {\it Comptonized} components. 

In particular,  D'Amico et al. (2001) studied only the high energy band above 20 keV using   
 High Energy X-Ray Timing Experiment (HEXTE) data 
and tried to find  a correlation between the 
X-ray  {\it hard~tail} 
emission in the spectra of Sco~X-1 
with the position of the source along {\it Z}-track. However, they did not find any 
apparent correlation 
using a {\it bremsstrahlung} component 
and a simple $power$-$law$ component in their spectral model.
Furthermore, \cite{D'Amico01} claimed a non-thermal 
origin of the hard tail. 
In turn, Bradshaw et al. (2003) used only  the Proportional Counter Array 
(PCA) data in the moderate energy range 3 -- 18 keV, applying a model consisted of a blackbody emission and a {\it Comptonization component} [BMC, \cite{ST98}]. They studied an evolution of X-ray spectrum of Sco~X-1 as a 
function of accretion rate and, in particular, detected an increase of the absorption at low energies a factor of 2 when the object moved from the HB to NB/FB vertex. 
 Barnard et al. (2003) used  the data from  the PCA and HEXTE, applying a {\it blackbody} component and a {\it cutoff 
power law} to the  data in the energy range 2.5 -- 50 keV. In the framework of this model, the so-called Birmingham
model [\cite{Church01}] 
they studied an evolution of the blackbody emission (from the NS) and Comptonized emission (from an extended accretion 
disk corona) between different spectral states.  \cite{Church12} also used data from  PCA and
HEXTE, showing that the spectrum, in the energy range 2 -- 50 keV, can be
approximated by a soft $blackbody$ component plus a {\it cutoff power law} and a broad $Gaussian$ line. As a result, they 
revealed   the apparent  non-monotonic behavior of  mass accretion rate in Sco~X-1 along 
{\it Z-}track. They also argued 
that the mass accretion rate increase is a function of 
the neutron star temperature.
\cite{D'Ai07} also analyzed the mass accretion rate variations  studying energy spectra from selected regions in the {\it Z-}track 
of the CD.
They demonstrated that X-ray spectra of Sco~X-1 based  on PCA \& HEXTE observations can be adequately fitted by 
a three-component model, consisting of a soft thermal component ($diskbb$), 
a thermal Comptonization component ($compps$/$comptt$/$thcomp$) and a power-law component ({\it pegpwrlw}).
Revnivtsev \& Gilfanov (2006) also studied the contribution of the boundary layer in the emergent X-ray spectrum of NS LMXBs  
as a function of the {\it Z}-track position. 
They fitted the spectra of various NS sources using two Comptonization ($CompTT$) components, one is related to the NS 
emission and another to the transition 
layer. 
 Gilfanov \& Revnvitev (2005) also argued  that the boundary layer is the source of rapid variability based on Fourier-Revolved spectral component analysis. 

In this Paper we present the analysis of the 
{\it RXTE} observations during  1996 -- 2002  for 
Sco~X-1.  In \S 2 we present the list of observations used in our data analysis.  
In \S 3 we provide results of our X-ray spectral analysis and in \S 4 we interpret the observed 
X-ray spectral properties using the transition layer model. 
We make our  final conclusions in  \S 5.   

 
\section{Data Reduction and Analysis \label{data}}

We  have 
used publicly available  {\it RXTE} data 
 from Sco X-1 obtained from May 1996 to May 2002. 
In total, these data  are derived from 60 observations taken at distinct spectral
states of the source.
Standard tasks of the LHEASOFT/FTOOLS
5.3 software package were utilized for data processing.
For spectral analysis we used PCA {\it Standard 2} mode data, collected 
in the 3 -- 23~keV energy range, using the most recent release of PCA response 
calibration (ftool pcarmf v11.7).  The standard dead time correction procedure 
has been applied to the data. 
We also used  data from the HEXTE detectors 
to construct  the background-subtracted  broad-band spectra.
Only data  in the 19 -- 250~keV energy range were used for the spectral analysis in order 
to account for the uncertainties of the HEXTE response and 
background determination. 
 The data are available through the GSFC public archive 
(http://heasarc.gsfc.nasa.gov). In  Table 1 we list the  groups
 of observations covering complete range of the source evolution 
 during different spectral  state events. 

We have thus   made an analysis of {\it RXTE} observations  of Sco~X-1  spanning seven years 
for six intervals indicated in Table 1. 
{
To  model the spectral evolution of the source we made spectra on an appropriate timescale. 
This timescale should be short enough to describe  a spectral evolution, 
but  be long enough to provide a realible statistics. We selected regions in the color-color diagrams
(CCDs) (see Figs.~\ref{HID_Sco} and \ref{CCD_Sco}) for each observational  data set and create the CCD-selected spectra.  
For our analysis we use version 6.0 of the FTOOLS package and version 12.1 of XSPEC. 
}
A systematic error of 0.5\% 
have been applied to {\it all}  analyzed spectra to account for the absolute calibration uncertainties.  
 We have also used  data from the  All-Sky Monitor (ASM) on-board {\it RXTE} to construct a long baseline 
3 -- 10 keV light curve bracketing the PCA and HEXTE  observations. 


ASM (2 -- 12 keV) light curve of Sco~X-1 demonstrates 
irregular flaring activity with the intervals of more frequent flares as well as  the intervals 
of reducing flaring. Thorough analysis of the flares detected in such a manner shows that 
the spacing of adjacent flares varies between 0.3 and 2.5 days in most cases.
Less frequently flares  have much larger
gaps about 10 days. Because of this 
we have  chosen  the observational sets consisting  
flaring patterns for our detailed analysis  of Sco~X-1.



\section{Results \label{results}}

\subsection{Color-color and hardness-intensity diagrams of Sco~X-1 \label{ccd}}

{
We investigate the light curves with 16s time binning for four energy 
channel ranges  5 -- 10, 11 -- 24, 25 -- 54 and 55 -- 107. These energy ranges correspond to  energies 
 1.94 -- 4.05 keV, 4.05 -- 9.03 keV, 9.03 -- 20.3 keV, and 20.3 -- 39.99 keV, respectively. 
We constructed 
CCDs 
of the source using the energy-dependent light curves and defined the soft color (SC) as a ratio 
of count rates in the 4.05 -- 9.03 keV and 1.94 -- 4.05 keV energy bands, while the hard color (HC)
is defined as a ratio of count rates in the 20.3 -- 39.99 keV and 9.03 -- 20.3 keV energy ranges. 
Note that above channel-to-energy conversion is given for epoch 3, which is  presented 
for most of our data ($R1$ -- $R5$; see Table 1). For the data set $R6$ the energy ranges are slightly shifted 
taking into account the change of the instrumental epoch of the satellite (epoch 5), which lead to  
a little 
shift on the HC and on the SC axes.

The obtained CCDs and HIDs {\it hardness intensity diagrams} 
are shown in Fig.~\ref{HID_Sco} for available {\it RXTE} data of Sco~X-1 
using bin size 16 s. They present a 
family of tracks, 
which are characterized by a clear $\nu$-like shape.  
The left part of $\nu$-like track is related to the {\it horizontal} (HB)
 and 
 {\it normal} 
(NB) branches, 
 while the right part of this $\nu$-like track corresponds 
to the {\it flaring} branch (FB). 
The sharp minimum of such a $\nu$-like track is 
the so called {\it soft~apex} (or NB/FB vertex),  
where the luminosity 
of Sco~X-1 notably decreases. 
 The typical error bars for the colors are shown in the bottom right corner of left panel; errors
in the intensity are negligible. 

As one can  see from this Figure the CCDs and HIDs 
provide a family of 
{\it Z-}cycles for all data  sets (see Table 1). It is worth noting that  these particular sets   partly 
overlap  each other. 
The sets with strong flaring activity are pointed out by {\it black} points, while the sets with 
reduced flaring activity  are highlighted by {\it red} points. 
 The  ``reduced flaring activity''  set shows underdeveloped {\it Z-}cycle. 
Three typical tracks with directions of HB$\to$NB$\to$FB transitions for strong flaring activity sets are indicated by corresponding bended arrows. 
The shifted {\it Z-}tracks  in CCD and HID 
can be seen  as a 
NB/FB vertex shift. 
The evolution of the NB/FB vertex position in Sco~X-1 cannot be described as one-to-one correspondence 
 with X-ray luminosity as in the case of  XTE~J1701-462, when the position of NB/FB vertex (at Sco-like stage)
moves with X-ray luminosity with a decrease  of SC and HC (see Fig.5 of LRH09).
In order to understand the X-ray evolution of 
 Sco~X-1 
one needs to investigate 
 the {\it Z-}track behavoir using
spectral (\S.~\ref{spectral analysis}) and timing analysis (\S~\ref{transitions}) together.  
}

\subsection{Evolution of spectral hardness throughout all {\it Z-}branches of Sco~X-1
\label{ccd}}

{
Evolution sequences along  the CCD and HID tracks of Sco~X-1 are presented in Figs.~\ref{CCD_Sco} and 
\ref{all_HID_Sco} 
in ten 
panels for corresponding subgroups of successive observational intervals. 
In particular,  in Fig.~\ref{CCD_Sco} we  indicate 
selected regions using different colors which are related to 
the CCD-resolved spectra.
We label the spectra through a progressive numbering (according to Table 2, 
first column). 

As seen from Fig.~\ref{CCD_Sco}, 
the CCD tracks 
are  extended,  
in particular, related to the FB (see II -- VII panels). 
It is worth noting that the source  moves back and force along {\it Z-}track for these particular observations.
We select regions in the CCDs for each observational  data set, marked with superscripts, e.g. ``a'', ``b'' etc. 
In contrast,  relative compact areas  are seen in I, VIII, IX, X panels.
During these observations 
the source evolves around a short {\it Z-}track segment and thus one can conclude that 
the  spectral variability is small for these observational segments.
Using these selection criteria 
 we eleborated the 
time intervals, 
 which we applied to extract 
the relevant spectra for the PCA and HEXTE data sets.

The {\it Z-}shape evolution and its shift 
can be  clearly 
seen in Fig.~\ref{all_HID_Sco}. 
The evolution of the HID tracks 
as a function of the count rate in (9 -- 20 keV) energy band 
is characterized by a significant HC variability 
(0.01$<$HC$<$0.03). One can see a well-developed FB for  high
intensities  (I -- VII groups), while  a moderate HC variability  (0.01$<$HC$<$0.016) is observed for  
relatively  lower
 intensities  (VIII -- X groups). 

Note that the FB in Sco~X-1 is
associated with a monotonic increase of the HC with respect to the  total intensity 
from the bottom of the flaring branch to the top  (botFB and topFB respectively). 
This behavior of the  HID  is in contrast to that in  the {\it dipping Z}-source GX~340+0, which  shows 
only a slight increase and sometimes even  a decrease of the HC with the total flux during the entire FB [see Hasinger \& van der Klis (1989);  Kuulkers \& van der Klis (1996) and Seifina et al. (2013), hereafter STF13].  

It is also very interesting to compare our study of CCD and HID  for  Sco~X-1 with those 
obtained by LRH09 for  XTE~J1701-462. Whereas the energy bands 
used for CCD and HID are different
in  LRH09 and our investigations  
for  Sco-like {\it Z-}tracks   their  evolutions demonstrates 
changes of the track shape and its positions are similar. 
In particular, LRH09 
found that 
the position of NB/FB vertex of {\it Z-}track for XTE~J1701-462 is shifted approximately along inclined line with positive and negative inclinations for CCD and HID, respectively. 
(e.g., see panels II and III of Fig. 5 in  LRH09). 
The shapes of {\it Z-}track for XTE~J1701-462 
evolves (in CCD and HID)  while 
its $\nu$-shape morphology remains,  having 
clear elements of the HB, NB and FB and 
well-determined NB/FB vertex. In contrast, 
the HC of NB/FB vertex in Sco~X-1 changes  
with X-ray intensity (compare $pink$ and $green$ line tracks in Fig.~\ref{HID_Sco}) as that in XTE~J1701-462, but then 
the HC increases with the intensity. 
}

\subsection{Spectral Analysis \label{spectral analysis}}


 {\it Z-}sources  vary on timescales of minutes to hours. 
This is a subject of many previous investigations of Sco X-1 (see, for example, van der Klis et al., 1996; Bradshaw et al., 2003; Barnard et al., 2003; Belloni et al., 2004). 
Here we  concentrate on 
specific properties of Sco~X-1 related to the hard X-ray emission. The source + background spectra have been compared 
with the background spectra for both the PCA and HEXTE in order to estimate the significance of the hard tail detection  
and to obtain the high energy component of the spectrum as best as possible. 
In addition, we excluded the intervals, which do not provide  significant signal for high energy range ($>$30 keV).  

As the first trial, we tested 
a model that consisted of an absorbed thermal component $Bbody$, a thermal Comptonization 
component {\it Comptb} (Farinelli et al. 2008) 
and a {\it Gaussian} line component.  But this model [$wabs*(Bbody+Comptb+Gauss$)] 
gave a poor description of about 
60\% of the data, in particular, of the FB spectra. Significant 
negative residuals at low energies (less than 4 keV) and greater than 
30 keV suggest
a presence of additional  emission components. 
Because of  this reason and following  suggestions by Farinelli et al.   we also attempted  a  
 {\it two-Comptb} model  $wabs*(Comptb+Comptb+Gauss)$. As a result of these efforts we have found  satisfactory  fits for all available sets of the  data 
(during all spectral states).

The {\it Comptb} model describes an emergent spectrum as a convolution 
of an input or {\it seed} black body spectrum having
the normalization $N_{com}$ and the  seed photon temperature $kT_s$
 with a Comptonization  Green function.
Similar to the ordinary {\it Bbody} XSPEC model,
the normalization $N_{com}$ is a ratio of the source 
luminosity
to square of the distance $D$
\begin{equation}
N_{com}=\biggl(\frac{L}{10^{39}\mathrm{erg/s}}\biggr)\biggl(\frac{10\,\mathrm{kpc}}{D}\biggr)^2.
\label{comptb_norm}
\end{equation}  

It is worthwhile to emphasize that the Comptb model is an updated version  of the COMPTT model of XPEC (see Titarchuk 1994) but with only difference that the main parameters in the former model are the seed photon temperature $kT_s$, the electron temperature $kT_e$, the spectral index  $\alpha$ and parameter $A$ related to the illumination fraction of the Compton cloud by soft (seed) photons $f=A/(1+A)$.  

{This model describes} a scenario 
in which a Keplerian disk  is connected to the NS 
through  the transition layer  (TL) 
[see  the description of this scenario in Titarchuk et al. (1998)].
In Figure~\ref{geometry}  we illustrate  our spectral model in application to Sco~X-1.
%
%
We assume that accretion into the NS 
takes place when the material passes through 
the two main regions:  a geometrically thin accretion disk [for example, standard Shakura-Sunyaev 
disk, see \citet{ss73}]
and the TL,
where the NS surface  and   soft disk photons   are  upscattered off hot electrons. 
In our picture 
the emergent thermal Comptonization spectrum is  formed in  the TL, 
where 
disk   photons of temperature $kT_{s1}$
 and soft photons from the NS photosphere
 of temperature $kT_{s2}$
 are scattered off the 
 TL hot electrons   giving rise to 
two Comptonized components, herein {\it Comptb1} and {\it Comptb2}, respectively.
Red and blue photon trajectories, shown in Fig.~\ref{geometry},  correspond to the soft 
(seed) and hard (upscattered) photons, respectively. 

In the framework  of the applied model $wabs*(Comptb1+Comptb2+Gauss)$, the free parameters of the model are:
$\alpha_1$, $\alpha_2$; 
$kT_{s1}$, $kT_{s2}$; $log(A_1)$, $\log(A_2)$ which are related to the Comptonized
fractions $f_1$, $f_2$;
$kT^{(1)}_e$ and  $kT^{(2)}_e$; the normalizations of the BB
components  $N_{Com1}$ and $N_{Com2}$ of the $Comptb1$ and $Comptb2$
respectively. 
%
We also add to our model  a {\it Gaussian} component, 
whose  parameters are  a centroid  energy $E_{line}$, the 
width of emission line  $\sigma_{line}$  
and normalization $N_{line}$.


It should be noted  that we  fixed certain parameters of the $Comptb$ models: 
$\gamma_{1/2}=3$, which are related to the index of the low energy part of the spectrum, namely   
$\alpha_{1/2}=\gamma_{1/2}-1=2$, 
and $\delta_{1/2}=0$ because we neglect an efficiency  of the  bulk inflow effect vs the  thermal Comptonization   
for  NS accretion.
The bulk inflow  Comptonization should take place very close to NS surface. However  if the 
radiation pressure in that vicinity is sufficiently high then the bulk motion is suppressed. On the other hand if mass accretion  is quite low then the effect of the bulk motion 
is negligible. Generally, the bulk effect in NSs is a  rare event. 
We  also use a fixed value of hydrogen column $N_H=3\times 10^{21}$ cm$^{-2}$, which was found by Christian \& Swank (1997). 
We also fixed the value of the $Comptb$ parameter $log(A_2)$ to 2 when the best-fit values of $\log(A_2)\gg1$ because
in any case of $\log(A_2)\gg1$ a Comptonization fraction $f = A/(1 + A)$ approaches unity 
and further variations of $A\gg1$ do not improve fit quality any more. 

Note that in a case of  the {\it two-Comptb} model, we deal with two seed-photon temperatures, in which a low value of the temperature  cannot be determined, because the PCA low-energy threshold ($\sim$~3 keV) is  above the peak of the disk seed-photon blackbody which
temperature, typically  $\lax$0.7 keV  
in LMXBs,
 [see Barret 2001, \cite{ST11}, \cite{ST12}, \cite{STF13}]. 
The results of these fits to the $two$-$Comptb$ model are reported in Table 2.

Generally, the adopted spectral model for Sco~X-1 exhibits a 
fidelity throughout all the data sets used in our analysis. Namely, a value of reduced
$\chi^2_{red}=\chi^2/N_{dof}$, where $N_{dof}$ is a number of degree of freedom, 
is  less than or about 1.0 for most observations. For a small 
fraction (less than 2\%) of spectra with high counting statistics
$\chi^2_{red}$ reaches 1.4. However, it never exceeds a rejection 
limit of 1.5 (for 90\% confidence level). 
Note that the energy range for the cases, in which we obtain  the  poor fit statistics 
(one among 98 
CCD-resolved spectra with $\chi^2$=1.46 for 87 d.o.f),   are related to the iron line region.  
The line  width 
$\sigma_{line}$ 
does not  vary much  and previous measurements  suggest that it is  in the range of 
0.7 -- 0.8 keV. Therefore, we fixed $\sigma_{line}$ of $Gaussian$ component at a value of 0.8 keV. 
In this approach we detected only a moderate contribution of the $Gaussian$ component into the spectrum 
throught all {\it Z} spectral states except for some ones
at NB/FB vertex region. 
{ 
It is worth noting that LRH09 also detected a weak iron line  emission line in the spectrum 
of  XTE~J1701-462 during all {\it Z-}states with some increase at the lower part of the NB. 
}

\subsection{Spectral properties as a function of {\it Z}-track stage \label{evolution}}

In Figure~\ref{sp_compar_xte}, on the {\it left}  panel  we present the HID where the branches are traced out 
by arrows  as
HB$\to$NB$\to$bottomFB 
(on the left part)   and then from the bottomFB to the  topFB  (on the right part). Here we also point out different regions of {\it Z-}track in HID by different colors.
In the  {\it central and right} panels  we show six representative $EF_E$ spectral diagrams  for the $left$ and 
$right$ parts of  $\nu$-shaped  {\it Z} track respectively.
In  the {\it central} panel the spectra correspond  to {\it RXTE} observations 
20053-01-01-03 ({\it green}, botNB) 
20053-01-01-06$^b$  
({\it pink}, NB, see Table 2)
and 
20053-01-01-00 ({\it violet},  HB).
On the {\it right}  panel we present three other spectral diagrams
 for the {\it right} part of  {\it Z}  track from (botFB$\to$topFB).
 The corresponding data are taken from   {\it RXTE} observations 20053-01-01-02$^e$ 
 ({\it bright~blue},  Table 2), 
botFB), 20053-01-02-01$^a$ ({\it red}, midFB,  Table 2) 
and  20053-01-02-04 ({\it black}, topFB). 
The corresponding data are taken from {\it RXTE} observations 
20053-01-01-02$^e$ ($bright~blue$, 
{botFB},  Table 2), 
20053-01-02-01$^a$ ({\it red}, midFB,  Table 2) 
and  20053-01-02-04$^{a-b}$ ({\it black}, topFB,  Table 2). 
Using this Figure one can see changes of  the spectral 
shape for the
{HB}, 
{NB}
and  
{FB} spectral branches. In particular, the hard tail gradually decreases when the source moves from the  HB
through the NB to the botNB.  However, the hard emission becomes stronger when the source evolves from 
the botNB to the topFB 
(see the left and right panels of Fig. \ref{sp_compar_xte}).   

In Figure~\ref{Zsp_compar_RXTE}  we  also show three representative $EF_E$ spectral diagrams for 
different states along the {\it Z}-track.
For these spectra data are taken from {\it RXTE} observations 20053-01-01-00
({HB})
20053-01-01-06$^b$
({NB}, Table 2) and 
20053-01-01-02$^e$
({FB}, Table 2). The data are shown by  black points and  
 the spectral model components are displayed  by {\it  blue, red} and 
{\it purple} lines for $Comptb1$, $Comptb2$, 
 $Blackbody$ and $Gaussian$ respectively.
 Yellow shaded areas demonstrate an evolution of
$Comptb1$   between the {HB},
{NB} 
and {FB} 
branches when $kT^{(1)}_e$ 
monotonically increases from 3 keV to 180 keV.
In this  Figure 
one can clearly see  changes of spectral shape in the energy range greater than 30 keV and the relative contribution 
of Comptonized components  for different   {\it Z-}branches. In particular, the {NB} panel demonstrates the  relative 
softening of the $Comptb1$ component 
in comparison of that presented in the {HB} panel. While the {FB} panel shows the hardening of {\it Comptb1} component 
in comparison of that seen in the {HB-NB} panels. We also detect a relative growth of the {\it Gaussian} iron line emission at the bottom of the NB  (as one can see from panel 
$Z2$ of Fig.~\ref{Zsp_compar_RXTE}). 

{
Spectral changes are also seen in Figs.~\ref{lc_r1} and \ref{lc_low}, which show 
strong and reduced flaring activity sets as a function of time.
In particular, the electron temperature $kT^{(1)}_e$ ($blue$ points)  ranges between 20 keV and 50 keV during the HB and NB 
states (see MJD 50556, 50561 -- 50562)
and then it decreases to 5 keV at the $soft$ apex (see
MJD 50559 in Fig.~\ref{lc_r1} and MJD 51341 in Fig.~\ref{lc_low}).  
$kT^{(1)}_e$ reaches its maximum at  200 keV  during the FB   (see MJD 50557, 50560 marked with blue vertical strips in Fig.~\ref{lc_r1}). On the other hand, the $Comptb$ normalizations $N_{Com1}$ and $N_{Com2}$ are only  weakly  correlated with  
the variations of  {\it flux ratio} coefficient FR and  
 count rate in the 2 -- 9 keV energy band ($black$ points), which is presumably   related to 
the NS emission that dominates during all spectral states (see Fig.~\ref{lc_r1}). 
 The normalization  $N_{Com2}$
is low variable from 0.4 to 3 in units of $L_{39}/{D^2}_{10}$ (where $L_{39}$ is the  luminosity of the seed blackbody 
component in units $10^{39}$ erg s$^{-1}$ and $D_{10}$ is distance in units 10 kpc). {The X-ray 
contribution from the Comptonizaton of the seed photons coming from the disk 
is weaker by factor 2 than that related to the seed photons coming from the NS surface},  
 during all spectral states (Fig.~\ref{lc_r1}). 
It is worth noting that the energy spectral indices  $\alpha$
  shown in the bottom panels of Figures \ref{lc_r1} -- \ref{lc_low} only  slightly vary with time 
about a mean value of one
except for 
index $\alpha_1$ at the FB intervals (marked by $blue$ vertical strips in Fig.~\ref{lc_r1}). 
At that time the index  $\alpha_1$ significantly 
decreases down to 0.3 [see  MJD 50557, 50560 in Fig.~\ref{lc_r1}].
These changes are also seen in Fig.~\ref{te1-Zstate} 
along the total {\it Z}-cycle of Sco~X-1. 
}

{ We can compare
our spectral results on Sco~X-1 with the spectral evolution of 
 XTE~J1701-462. 
In particular, LRH09 used a spectral model which was consist of BB, MCD (multicolor disk), $Gaussian$ and  a  CBPL (Comptonized broken power law) 
components to fit 
the spectrum of XTE~J1701-462.
In this model they 
found that  the spectrum of XTE~J1701-462 on the HB of {\it Z-}stage 
shows 
a hard tail (see 
Fig.~13 in LRH09) 
which can be explained  by Comptonization of the soft  photons originated  in the innermost  disk region. 
We obtain a similar result 
for the Sco~X-1 case.  Namely, using our   Comptonization component ($Comptb1$), 
indicated by the yellow shaded 
area in  Fig.~\ref{Zsp_compar_RXTE} (left panel), we reveal  a hard  emission in the HB stage. 
Furthermore, LRH09  found that   
the spectra of XTE~J1701-462 corresponding to the HB/NB vertex, NB and NB/FB vertex 
did not have  
the Comptonized component (in their terms, CBPL component). 
In Sco~X-1 we detect the Comptonized component ($Comptb1$) at the NB but it  is  rather weak 
 (see the central panel of  Fig.~\ref{Zsp_compar_RXTE}).
In the FB  the spectrum of XTE~J1701-462 was mainly characterized by BB+MCD components and with  some indication to  
a weak hard emission (see Fig.~13 in LRH09).  
In contrast 
to LRH09, we detect the strong  Comptonized component ($Comptb1$) at the FB  of Sco~X-1 spectrum up to 250 keV (see the right panel of Fig. \ref{Zsp_compar_RXTE}). Furthermore, 
the photon index of $Comptb1$ is surprisingly decreased during the  FB, which is an indication of  the hardening of X-ray  
emission in Sco~X-1. It is  interesting to note that another {\it Z-}source GX~340+0 
does not show a significant Comptonized component at the FB state (see STF13). Such a difference can be related to different distances to these sources 
($D_{ScoX-1}=2.8$ kpc, $D_{1701}=8.8$ kpc, $D_{GX340+0}=11$ kpc). Thus the proximity of Sco~X-1 to the Earth   provides us  a unique opportunity to  study  the FB stage for {\it Z-}sources. 
}

\subsection{{\it Z-}state evolution of X-ray spectral parameters in Sco~X-1 as a function of the electron temperature $kT_e^{(1)}$ \label{energy spectra}}

{
}

In Figure~\ref{te1-Zstate}  we present the dependencies of the photon index $\Gamma$, of the $Comptb$ normalization $N_{com}$, 
of the illumination fraction $f$, of the flux ratio and of the X-ray flux on the electron temperature 
$kT_e^{(1)}$. Here we mark the HB and NB states by a $violet$ and $blue$ 
vertical strips, respectively. 
In addition, the $hard$ and $soft$ apexes are indicated by arrows and a $pink$ vertical 
strip highlights the interval of mid-topFB stage where the photon index of the $hard$ component $\Gamma_1$  significantly decreases (from $2.0$ to $1.3$). As it is clearly seen from this Figure 
the FB, NB and HB stages of {\it Z-}cycle can be  related to a change of $kT_e^{(1)}$ (in the framework  of our model). 

\subsubsection{The FB  \label{FB}}

The FB of the Sco~X-1 spectrum is characterized by a wide range of the electron temperature $kT_e^{(1)}$ 
(from 5  to 200 keV). 
High and low values of the $kT_e^{(1)}$ occur  at the topFB and  near $soft$ apex respectively (see hardness-intensity diagram on the left part of Fig. \ref{sp_compar_xte}).  
The photon index $\Gamma_1$  shows a specific behavior during the FB. 
In Fig. \ref{te1-Zstate}  (see panel $a$, $blue$ points) one can see that  $\Gamma_1$  decreases from 2 to 1.3 
when the  total X-ray flux increases (see panel $e$). 
In contrast, the $soft$ Comptonized component of X-ray spectrum (Com 2) 
shows 
that $\Gamma_2$ only slightly varies around 2 (see red points in panel $a$). 
When the object goes into  the midFB [see $pink$ vertical strip in panel $b$ of Fig.~\ref{te1-Zstate}]
the COMPTB normalization  $N_{com1}$  drops by a factor of 4 -- 5, reaching  the lowest level 
($N_{com1}<0.25~L_{39}/D^2_{10}$, panel $b$, $blue$ points).  On the other hand  $N_{com2}$  
monotonically  increases and exceeds  $3~L_{39}/D^2_{10}$ (see panel $b$, $red$ points).  
At this stage we find that 
$\Gamma_1$ and $N_{com1}$ of the $hard$ Comptonized component drop.
In the FB the illumination fractions of both 
components $f_1$ and $f_2$  are quite different. While  values of $f_1$ are low ($f_1<0.1$, $blue$ points), 
those of $f_2$ is high ($f_2>0.6$, $red$ points of panel $c$). 
The mid-topFB spectra  are characterized by a high hardness ratio  (see panel $d$) and high level of
 the 3 -- 250 keV flux ($> 3\times 10^{-7}$ erg/s/cm$^2$, see panel $e$).    
Note also that the source behavior during the FB was evaluated using  CCD-resolved and HID intervals, shown in Fig.~\ref{CCD_Sco} and \ref{all_HID_Sco}.  
 Different {\it Z} cycles  belong to  different areas in CCD and HID although 
the FB spectral properies for them are similar and almost independent of  {\it Z} cycle change.





\subsubsection{The NB/FB vertex  \label{NB_FB}}


The NB/FB vertex of {\it Z-}track for Sco X-1 (indicated as ``$soft$ apex'' in Figure~\ref{sp_compar_xte})
is associated 
with 
the lowest value of the electron temperature $kT_e^{(1)}$ ($kT^{(1)}_e\sim 3$ keV). This apex is a boundary between 
two phases for the photon index behavior,  $\Gamma_1=2$ (at the NB-HB) and $\Gamma_1<2$ (at the FB). 
Panel $a$ of  Fig.~\ref{te1-Zstate} demonstrates that at this apex 
$\Gamma_1$ ($blue$ points) and $\Gamma_2$ ($red$ points) are both equal to 2. Moreover,  the $soft$ apex (NB/FB vertex) is characterized by sharp changes of the $Comptb$ normalizations and the illumination factors 
for both Comptonized components $f_1$ and $f_2$  as variations of $kT_e^{(1)}$  are small (see panels $b$ and $c$ of  Fig.~\ref{te1-Zstate}). As one can see  from this Figure that $f_1$ and $f_2$ anticorrelate with each other when $kT_e^{1}$ increases.  
For the $hard$ Comptonized component  {\it Com 1}   the normalization $N_{com1}$ decreases from  
$2\times L_{39}/D^2_{10}$ to $0.2\times L_{39}/D^2_{10}$ and $f_1$ decreases from 0.6 t 0.05   when $kT_e^{(1)}$ increases from 3 to 40 keV (see panels $b$ and $c$ respectively). 
While  the normalization of the $soft$ Comptonized component $N_{com2}$ increases from $0.5\times L_{39}/D^2_{10}$ to $2.5\times L_{39}/D^2_{10}$ a fraction  $f_2$ increases and saturates at unity (see panels $b$ and $c$ respectively).
In the NB/FB vertex ({\it soft} apex) the spectra of Sco~X-1 are characterized by
lower values of flux ratio 
and  the 3 -- 200 keV flux 
(see panels $d$ and $e$ respectively).

\subsubsection{The NB  \label{NB}}


The NB of {\it Z-}track for Sco X-1 (indicated by $bright~blue$ vertical strip in Fig.~\ref{te1-Zstate}) is associated 
with 
a specific range of the electron temperature $kT_e^{(1)}$.  When 
the source transits from HB/NB vertex ($hard$ apex) to NB/FB vertex ($soft$ apex,  
 see left panel of Fig. \ref{sp_compar_xte})
$kT_e^{(1)}$ monotonically decreases 
from 40 keV to 5 keV. 
The photon indices $\Gamma_1$, $\Gamma_2$ throughout NB state are  around 2 (see $blue$ and $red$ points of panel $a$ in 
Figure~\ref{te1-Zstate}). 
The COMPTB normalization parameters ($N_{com1,2}$) 
monotonicaly increase by a factor of 5 during the NB (see panel $b$ of Fig.~\ref{te1-Zstate}). 
The {\it hard} Comptonized component  ({\it Com 1}) is characterized by a significant illumination fraction $f_1$ only 
in the vicinity of $soft$ apex, but then throughout 
the rest of the NB $f_1$ is extremely low. While   
the illumination fraction $f_2$ 
related to the second Comptonization component  
({\it Com 2}) 
is   high ($f_2\to 1$) over the NB (see the left bottom panel of Fig. \ref{te1-Zstate}). 
The NB spectra of Sco~X-1 are characterized by a variable flux ratio (see panel $d$) and a wide range of the 
3 -- 200 keV flux [$(1.5-3.6)\times 10^{-7}$ erg/s/cm$^2$, panel $e$]. Note also that  
in the NB the spectrum  is relatively  soft  (see middle panel of Fig. \ref{Zsp_compar_RXTE}). 

\subsubsection{The HB  \label{HB}}


The HB of {\it Z-}track, indicated by $violet$ vertical strip in Fig.~\ref{te1-Zstate},  is associated 
with 
a small range of the electron temperature $kT_e^{(1)}$ ($40<kT^{(1)}_e<50$ keV). The photon indices $\Gamma_1$, 
$\Gamma_2$ throughout the HB  are  around 2 ($blue$ and $red$ points of panel $a$ in 
Figure~\ref{te1-Zstate}). 
The COMPTB normalization of $hard$ component $N_{com1}$ remains at the level of $1.5\times L_{39}/D^2_{10}$ ($blue$ 
points in panel $b$), while the $soft$ component is   dominant [see $red$ points, $2<N_{com2}<2.7~L_{39}/D^2_{10}$].
In the HB  the fraction  
 $f_2$ is variable in a wide range (from 0.6 to 1), while the fraction $f_1$ is  extremely low 
($f_1<0.1$, see panel $c$).
In the HB the spectra  are characterized by a high flux ratio (see panel $d$) and a relatively high  
3 -- 200 keV flux [$(1.5\div 3.6)\times 10^{-7}$ erg/s/cm$^2$, see panel $e$]. Note also that in this 
state 
the spectrum has a  hard tail 
characterized by 
$\Gamma_1 \approx 2$,  in contrast to the FB hard tail, for which   $\Gamma_1$ is in the range  
from 1.4 to 2. 


Thus, 
our analysis  allows us  to separate the contributions of the  two zones related to the {\it hard} and {\it soft} 
Comptonized components, one is related to the transition 
layer and another to the NS emission.
Furthermore, we 
detect a  significant increase of the $kT^{(1)}_e$ in the FB, in contrast to other {\it atoll} and {\it Z} sources 
in which the electron temperature  $kT_e$ usually  reduces during $flaring$ 
events [see e.g. \cite{ST12}, Church et al. (2014)]. 

  
\subsection{Timing properties of Sco~X-1 during NB-HB-FB evolution \label{transitions}}

The timing properties of Sco X-1 have been extensively studied previously and well documented by many authors 
[e.g., \citet{vdKl94}]. 
 However, given our new and  unique approach to model the spectral evolution of Sco~X-1 we have chosen to perform a joint spectral-temporal 
analysis for CCD-selected intervals  to search for deeper physical insight.
%
%
The {\it RXTE} light curves have been analyzed using the {\it powspec} task from
FTOOLS 5.1. The timing analysis {\it RXTE}/PCA data was performed in  2 -- 13 keV energy range 
using  the {\it binned} mode.  The time resolution for this mode is 1/128 s. 
 We found that the shape of the power density spectrum (PDS) at high frequencies is 
dominated by dead-time  effects (see also Zhang et al. 1995) which, in the case of Sco X-1 at a count rate 
$>$25 000 counts/s/PCU,  are large, uncertain and  not well understood to predict  a real shape of the 
PDS  high-frequency part. Because of this we use PDSs only up to about 100 Hz (see Fig. \ref{PDS}). 
The normalization parameter of {\it Powspec} was set to -2, such that Poisson (white) noise is subtracted 
and the remaining power integrates to give the excess variance in the lightcurve.
To investigate  an evolution of the source timing properties and use it for {\it Z-}branch identification
 we modeled the PDSs  using  analytic models and the $\chi^2$ minimization technique in the framework 
of QDP/PLT plotting package\footnote{http://heasarc.gsfc.nasa.gov/ftools/others/qdp/qdp.html}.

From the lower NB to the FB   Sco~X-1 shows  a monotonic increase of the very low frequency noise below 1 Hz (VLFN). The shape of a high frequency noise (HFN) component  can be  fit  by the exponentially cut-off power  law 
($\nu^{-\alpha_{HFN}} e^{-\nu/\nu_{cut}}$).
The HFN rms amplitude, in contrast to VLFN,  decreases along the {\it Z-}track (from $\sim$7\% to 3\%).

In addition to the noise 
continuum we found  in the PDSs  quasi-periodic oscillations (QPOs) between 5 Hz and 20 Hz, that can be modeled 
by Lorentzians.  In particular, we detected reliable
QPOs with centroid frequency  near 6 Hz when the source moves along
the NB.   We  also found  a continuous   increase of  QPO frequency from 6 Hz to 15 Hz when the source moves from the NB to the FB.  Then, during the FB   QPO frequency steeply increases from 
 10 Hz to 20 Hz  and finally disappears   at the highest count rates.
Note that almost the same timing behavior  was detected in another observation 
with EXOSAT [Predhorsky et al., 1986;  \cite{Dieters00}] and 
with {\it RXTE} [\cite{vanderKlis96}, \cite{Casella06}].
One can  relate  these QPO frequency changes 
over the FB  and its absence at the $top$ of the FB with significant changes of the energy 
spectra of Sco~X-1 (see \S~\ref{correlation}). Hereafter we use these timing signatures along with spectral 
signatures (\S~\ref{evolution}) for the  additional identification of {\it Z-}branch of Sco~X-1. 

\subsection{Correlation between  timing and spectral  properties during spectral state transitions \label{correlation}}

We  consider
the main properties of the power spectra along with those of the energy spectra. 
We find  a specific behavior of $\Gamma_1$ and the low QPO 
 changes during the  FB. Namely, we detect a significant change  of $\Gamma_1$  at the FB 
from quasi-constancy around $\Gamma_1=2$ that is established  for the HB-NB phases. To investigate  
the timing changes during this index change ($1.3<\Gamma_1<2$) we  present the results  of the FB state  
analysis  for  a time interval of  MJD=50556$-$50821  ($R2$ set). 
On the top of Fig.~\ref{PDS}, ($top~left$ panel) we show the  HID  corresponding to this time interval,  
as $red$/$blue$/$black$ points A, B, and C mark moments at 
MJD = 50557.2/50556.6/50819.6$^a$, 50818.7$^a$/50817.8/50558.7$^a$ and 50816.9$^c$/50817.8$^b$/50820.9$^e$ 
related to different phases of {\it Z}-branches. Here supersripts indicate
 the corresponding CCD-resolved intervals (in accordance to Table 2). 

As already mentioned above the {\it hardness  intensity diagram} (HID) 
of Sco~X-1 consists of two branches of $\nu$-shaped {\it Z-}track. The power and energy spectra of the left branch 
(HB-NB-Soft apex)  are presented at the bottom panels A in Fig. \ref{PDS}, while that of the right branch 
(soft apex-botFB-midFB-topFB) are shown  at two bottom panels B and C. The FB can be divided into three segments: $lower$ (B $blue$, B $red$), 
$middle$ (B $black$, C $red$) and $upper$  (C $blue$, C $black$). 
The FB 
(lower part) stage  is associated with 
the canonical $soft~apex$ along 
the {\it Z}-track.  As  we argue above  (see \S \ref{evolution}) we do not find significant 
differences of the energy spectra  between the $bottom$ NB point (A $black$) and the $bottom$ FB points (B $red$/$blue$). While the time variability properties 
for these points in the HID  are very different  (for comparison see the bottom left and right panels of Fig. \ref{PDS}). 

The   NS seed photon temperature $T_{s2}$  is about 1.5 keV when Sco~X-1 is around  
the botNB and the botFB (see 
Fig.~\ref{norm2-Zstate} for  $T_{s2}$ values). 
On the other hand a significant decrease of $T_{s2}$  from 1.5 to 0.7 keV 
was detected when the source transits from the  {\it soft} to {\it hard~apex} and from the midFB to the topFB states.
This significant change  of $T_{s2}$  in  the FB 
is a clear  indication of the NS photospheric expansion  due to  high radiation pressure from the  NS 
during this phase.
 
On the $bottom$ panels  of Fig.~\ref{PDS} we show PDS for 3 -- 13 keV  band ($left$ column), which are presented along   with spectra  [plotted  as $E*F(E)$] fitted to our model  ({\it right} column) for A ($red$), B ($blue$) and all C 
points  of  the  HID  shown on the left upper panel.
The strong  and broad the Normal Branch Oscillations (NBOs)  are seen at 
2$-$10 Hz [peaked at $\sim$6 Hz, FWHM=11.2$\pm$4.1 Hz, rms=5.3$\pm$1.0\% (A $red$) 
and FWHM=9.7$\pm$3.5 Hz, rms=4.8$\pm$1.2\% (A $blue$), respectively] during 
the  NB state. 
For the $bottom$ of the FB  (B $blue$, B $red$, see left upper panel)   the related PDS (see left bottom panel B1) shows two QPO peaks at 
10 Hz and 7 Hz  respectively  when the source moves from the bottom FB to the  middle FB.
On the other hand PDSs of  the $middle$ FB (B $black$, C $red$ points in the upper left panel) show  broad flaring branch oscillations (FBOs) 
around $\sim$7 Hz (FWHM=12.2$\pm$3.4 Hz, rms=7.5$\pm$1.4\%) 
along  with VLFN  ($\sim$5 \% rms).  As Sco~X-1 moves from the $middle$ to the $top$ of the FB we can clearly see an  increase of the QPO frequency from 6 Hz to at most $\sim$20 Hz (FWHM=10.2$\pm$2.5 Hz, rms=5.3$\pm$1.2\% for point C $blue$). The highest  QPO frequency measured ($\sim$21 Hz) was found  during a flare in January 1998 data (C $blue$ point). Finally,  on the $top$ of the FB 
(C $black$) we do not  find any QPO feature but only  the strengthened VLFN component ($\sim$3 \% rms).

Note that the similar QPO frequency evolution  between 8 Hz to 5 Hz when the source goes from the botFB to the midFB  was previously detected with EXOSAT by \cite{Dieters00}, however they did not detect any energy spectral changes. In contrast,  due to the broad-band {\it RXTE} coverage we discover significant changes of energy spectra 
at high energy range ($E>30$ keV) along with this $\nu_{QPO}$  {\it excursions}. 

 On the {\it right bottom} panels of Fig.~\ref{PDS} we present  spectra 
 for 3$-$250 keV energy range (panels A2, B2, C2),  
related to the corresponding  PDSs (see panels A1, B1, C1). 
 The data are shown by black points and  
 the spectral model components are displayed by dashed $red$, $blue$ and $purple$ lines for the $Comptb1$, $Comptb2$ and $Gaussian$ components respectively.  Yellow shaded areas demonstrate an evolution of the $Comptb1$ component during the state transition. In particular, 
the energy spectra for the groups  A and B  
are very similar,  while for group C  they show significant changes at high energy band ($E>30$ keV). 
For C $red$/$blue$/$black$ points on the left upper panel   we present all corresponding energy spectra  on the low right panel to demonstrate the spectral evolution for  the  $hard$ and $soft$ Comptonized components. 
The comparison of  the energy spectra in B2 and C2 panels  reveals 
 a noticeable decrease of $\Gamma_1$ photon index  for  the  group C. 
Furthermore, one can see that PDS  in the C case is completely different of that seen 
 in the B case.  

{
}

\section{Interpretation of the observed spectral properties}
\label{index reduction}

{
Before to proceed with the interpretation of the observations, let us briefly summarize them as follows. 
(1) The spectral data of Sco~X-1 are well fit by two {\it soft} and {\it hard} 
Comptonized ({\it Comptb}) components 
for all analyzed HB, NB and FB spectra.
(2) The {\it soft} Comptonized component is dominant 
throughout all spectral states 
(see Table 2, and  two normalization panels in Figures~\ref{lc_r1} and \ref{lc_low}).
(3) The high energy tail  of the spectra X-ray can be fit by the {\it hard} Comptonized component.
}

We find  a very unique drop of  $\Gamma_1$ when Sco~X-1 is in the FB and  the luminosity of Sco~X-1 is close 
to the critical value $L_{Edd}$. 
This fact is very surprising with  respect to other NS LMXBs [e.g., 4U~1728-34 (ST11), 4U~1820-30 (TSF13), 
GX~3+1 (ST12),  GX~340+0 (STF13)], which all demonstrate the  constancy of the index ($\Gamma_1\sim 2$). However, 
it is important to note that all  these sources, besides Sco X-1,   were observed at  a  sub-critical regime  
throughout all spectral states, while  in Sco~X-1 we find  the source  is at the Eddington regime. 
To take into account this peculiarity,  we explore the situation when the luminosity 
reaches its  critical value, which is applicable to the FB of Sco~X-1 case.

\subsection{Energy release in the NS transition layer and the spectral index of the emergent spectrum}

As \cite{ft11}, hereafter FT11, pointed out  the energy release in the transition layer (TL)  determines the spectral index of the emergent spectrum.  Namely,  FT11 [see also \cite{zs69}] demonstrated that the energy flux  per unit surface area of the TL (corona) can be found as 
\begin{equation}
Q_{cor}=20.2\int_0^{\tau_0}\varepsilon(\tau)T_e(\tau)d\tau,
\label{energy_release_ in_TL}
\end{equation}
where $T_e(\tau)$, $\varepsilon(\tau)$ and $\tau_0$  are the plasma (electron) temperature, the radiation density  distributions in the TL and its Thomson optical depth   respectively.   

We obtain the energy distribution $\varepsilon(\tau)$ as a solution of the diffusion equation 
\begin{equation}
\frac{d^2\varepsilon}{d\tau^2}=-\frac{Q_{cor}}{c}q(\tau)
\label{energy_distribution_ in_TL}
\end{equation}
where $c$ is the speed of light;  $q(\tau)=1/\tau_{\star}$ for $\tau\le\tau_{\star}$ and $q(\tau)=0$ for $\tau_{\star}\le\tau\le \tau_0$.
If one compares this equation with that similar in FT11 he/she finds that here we assume that the gravitational energy release takes please only in the part of the TL, $0\le\tau\le \tau_\star$ because the high radiation  (plus magnetic) pressure  from the NS  finally stops the accretion flow. 
We should also add the two boundary conditions at the inner TL boundary (which can be a surface of NS) and the outer boundary $\tau=0$.  They are correspondingly:
\begin{equation}
\frac{d\varepsilon}{d\tau}|_{\tau=\tau_0}=0, 
\label{inner_bound_cond}
\end{equation}
\begin{equation}
\frac{d\varepsilon}{d\tau}-\frac{3}{2}\varepsilon |_{\tau=0}=0. 
\label{outer_bound_cond}
\end{equation}

Using the expression for $q(\tau)$ (see Eq. \ref{energy_distribution_ in_TL} and below) we find  the solution 
$\varepsilon(\tau)$ of equations (\ref{energy_distribution_ in_TL}-\ref{outer_bound_cond})
\begin{equation}
\varepsilon(\tau)= \frac{3Q_{cor}}{c}[(\tau+2/3) - \tau^2/2\tau_{\star}]~~~{\rm for} ~~~0\le\tau\le\tau_{\star}
\label{mod_epsilon_tau_1}
\end{equation}
and 
\begin{equation} 
\varepsilon(\tau)= \frac{3Q_{cor}}{c}(\tau_{\star}/2+2/3) ~~~{\rm for} ~~~\tau_{\star}\le\tau
\le\tau_0.
\label{mod_epsilon_tau_2}
\end{equation}
Thus integration of $\varepsilon(\tau)$  gives us
\begin{equation}
\int_0^{\tau_0}\varepsilon(\tau)d\tau= \frac{Q_{cor}}{c}(\tau^2_{\star}+2\tau_0). 
\label{int_epsilon}
\end{equation}
It is easy to see that we obtain the identical result  with that in FT11 for $\tau_{\star}=\tau_0$. 
Now we can estimate the Comptonization Y-parameter in the transition layer (TL)  using
 Eqs. (\ref{energy_release_ in_TL}) and  (\ref{int_epsilon}). 
We rewrite Eq. (\ref{energy_release_ in_TL}) using the mean value theorem as
\begin{equation}
Q_{cor}=20.2\hat{T}_e \int_0^{\tau_0}\varepsilon(\tau)d\tau
\label{energy_release_ in_TL_m}
\end{equation}
where $\hat{T}_e$ is the mean temperature in the TL.
 Substitution of formula (\ref{int_epsilon}) in Eq. (\ref{energy_release_ in_TL_m}) 
\begin{equation}
\frac{k\hat{T}_e(\tau_{\star}^2+2\tau_0)}{m_ec^2}=0.25
\label{y_parameter_estimate1}
\end{equation} 
leading to the estimate of Y-parameter  in the TL [see a definition of  Y-parameter in 
\cite{rl79}]
\begin{equation}
Y=\frac{4k\hat{T}_e\tau_0(\tau_{0}+2)}{m_ec^2}=\frac{4\tau_{0}(\tau_{0}+2)}{\tau_{\star}^2+2\tau_0}
\label{y_parameter_estimate2}
\end{equation}

Now we use this formula for Y-parameter  for the determination of the spectral index $\alpha_{diff}$.
Namely we can write (see FT11)
\begin{equation}
\alpha =-\frac{3}{2}+\sqrt{\frac{9}{4}+Y^{-1}}.
\label{alpha_y_parameter}
\end{equation}
We rewrite the asymptotic form of this equation when $Y^{-1}\le1$ as
\begin{equation}
\alpha \approx \frac{Y^{-1}}{3}=\frac{4}{3}\frac{(\tau_\star/\tau_0)^2+2/\tau_0}{1+2/\tau_0}.
\label{asymp_alpha}
\end{equation}
 Thus one can see that the spectral index $\alpha< 1$ when $\tau_\star<\tau_0$ and 
 $\tau_0\gg 1$. 
In Figure \ref{PDS} (the upper right panel) we present the observed dependences of  the photon spectral index of the Comptonized component $\Gamma_1$ ($\alpha_1=\Gamma_1-1$)  and 
thus one can see that  $\Gamma_1$ is indeed  significantly less than $2$ (or $\alpha_1<1$) when the observed  luminosity  $L_{com2}$ coming from the TL inner part is around 1 and higher. This leads us to 
suggest that in this case the gravitational energy release takes place only  in the TL outer region where 
$L< L_{crit}$.   
   
 Now we show that  the critical luminosity $L_{crit}$, at which the radiation force equals to the gravitational force, higher than Eddington one for the electron temperature $kT_e\gax 50$ keV.

\subsection{Temperature correction of scattering cross-section for critical luminosity calculation}

In order to explain   the observed dependencies   of  $\Gamma$  vs   $L$  and $L$ vs $kT_e$ (see Fig. \ref{PDS} central and right upper panels) we have to estimate the critical luminosity  
for a given plasma characteristic  $kT_e$ and its presumed   spectrum $J_\nu(E)$.  To do this 
 we should calculate the weighted electron cross section  $<\sigma>$, using relativistic 
photon-Maxwellian electron cross section~\cite{Wienke85} for the normal chemical abundance 
\begin{equation}
<\sigma>=\frac{\int J_\nu (E) \sigma(E, kT_e)dE}{\int  J_\nu (E)dE},
\label{sigma}
\end{equation}
Then the critical  luminosity (which is the Eddington luminosity when  $<\sigma>=
\sigma_{\rm T}$)   for a given electron temperature one can find using an equality of the gravitational force and the radiation force  
\begin{equation}
 L_{crit}=\int  J_{\nu} (E)dE= \frac{4\pi c GM_{NS}}{<\sigma>}. 
\label{L_crit}
\end{equation}
We calculate $L_{crit}$ for  $M_{NS}=1.4 M_{\odot}$ and then
  we make a plot  $L_{crit}$ vs $kT_e$ (see solid $blue$ line in the central upper panel) and we see that this theoretical dependence passes through all observed data points  ($L$,  $kT_e$).

We  also show  here the level of   
the Eddington luminosity, assuming  that $\sigma=\sigma_{\rm T}$ (see a dashed $green$ horizontal line in the upper  central panel of  Fig. \ref{PDS}). 
It is  clear to see  from this plot that the critical luminosity $L_{crit} (T_e)$ is higher than $L_{Edd}$ for higher 
temperature $kT_e$ and  the  electron temperature correction of the scattering cross-section  explains an increase of $L_{crit}$ vs $T_e$.

\subsection{Comparison between  spectral 
characteristics of {\it  Z}-sources 
and {\it atoll} sources}

Now we know the correlations between spectral, timing properties  for a variable mass accretion rate observed in X-rays in  Sco X-1  during the evolution 
across the HB-NB-FB track. We  identify a new kind of the spectral index behavior in this source
in relation to other known NS LMXBs. For this reason, it is 
interesting to compare Sco~X-1 with other NSs and find  similarities  and differences between them.
Therefore   we present a comparative analysis for five
sources: {\it Z-}sources Sco~X-1, GX~340+0  (STF13)  and  {\it atoll}  sources GX~3+1 (ST12), 
4U~1820-30  (TSF13),  4U~1728-34 (ST11) using  the same spectral 
model which consists of the {\it Comptonized} continuum and the {\it Gaussian} line components. It is also interesting to 
compare our findings on Sco~X-1 with spectral behavior of the unique NS source XTE~J1701-462 since both objects show spectral 
transitions in  the super Eddington luminosity regime.  

\subsubsection{Quasi-constancy of the photon index at sub-Eddington regime and a  reduction  of the photon index $\Gamma_1$ at the  high luminosity state}

{\it Z-}sources Sco~X-1, GX~340+0 and $atoll$ sources, GX~3+1,  4U~1820-30, 4U~1728-34 demonstrate a similar behavior  of $\Gamma$ vs  $T_e$ in the {\it sub-Eddington} regime,  namely  the quasi-constancy of 
spectral  index $\Gamma$ which is 
distributed around 2. {\it It is  important  to emphasize  that we consider here the spectral behavior, which is related to the Comptonization region  which in our scenario is   the transition layer (TL).} 
According to FT11, ST11, ST12 and STF13, this result  can probably indicate that the gravitational energy release takes place in  the whole  TL,  namely $\tau_{\star}=\tau_0$    for these five sources and that is  much higher than the cooling flow of the soft disk photons  (see Eqs. \ref{y_parameter_estimate2}-\ref{alpha_y_parameter}). 

However, for  Sco~X-1  we test  a   wider range of the  luminosity 
than it  in other {\it Z} and {\it atoll} sources (see above).   In particular, Sco~X-1 allows us 
to establish the index behavior  at the critical  luminosity regime,  when the emergent luminosity is even higher 
than the Eddington luminosity.  We find 
that the photon index of  the TL region    $\Gamma_1$ significantly  decreases when  the source accretes  close to the  Eddington regime, which takes place in the FB  
(see Figs. \ref{PDS}, \ref{gam_te_5obj}). 
Namely, in this mid-topFB stage,
the $hard$ Comptonization tail is characterized by the reduced value of the photon index ($1.3<\Gamma_1<2$) and the high  electron temperature 
of the TL  ($60$ keV $<kT_e<200$ keV).
Despite  its low level of  normalization $N_{com1}$  the $hard$ Comptonized component $Comptb_1$  is  firmly detected  at least up to 200 keV in the FB.  Moreover the  fraction  of  the high energy emission  increases 
when $kT_e$ increases and  $\Gamma_1$ drops from 2 to 1.3.  It should 
 be emphasized  that the only Sco~X-1 among other {\it Z-}sources achieves  the critical luminosity regime 
(see the dark blue points in Fig. \ref{gam_te_5obj}). 
In turn, it is reasonably to compare this result on $atoll$ and {\it Z} sources with spectral properties of a unique source 
XTE~J1701-462 (LRH09) involving both $atoll$ and {\it Z} stages of evolution as well as reaching the Eddington luminosity. 
Whereas LHR09 apply 
different models to spectral fit of XTE~J1701-462 emission, their model 
assumes the role of weak Comptonization by the so called CBPL component (i.e. 
a constrained BPL with $E_b=20$ keV and $\Gamma\le 2.5$) in $soft$ states and by BPL in $hard$ states. In the frame of that model the observations of XTE~J1701-462 
in $hard$ states are dominated by the BPL component 
with 
photon index $\Gamma\sim 2$, while in $soft$ states the 
photon index often reaches its hard limit $\Gamma\sim 2$ and ranged up to $\Gamma\le 2.5$ in the spectral fit with the 
CBPL model, i.e.  $\Gamma$ is normally greater than 2. 
This disrepancy with our results for  $atoll$ and {\it Z} sources 
(for which $\Gamma\le 2$ in our model) can be related to different model 
composition and to different  energy range used for spectral fit. In fact, LRH09 analyzed XTE~J1701-462 spectrum from 3 keV only up to 80 keV. 

\subsubsection{Spectral hardness  and electron temperature variability  for {\it atoll} and  {\it Z-}sources:
comparison with    XTE~J1701-462\label{ccd}}

In order to compare evolution 
properties of Sco~X-1 and  GX~340+0 with {\it atolls} and BHC sources 
we plot in Fig.~\ref{HID_7object} (upper panel)  HC (10-50 keV/3-50 keV) versus the luminosity 
in the 3 -- 10 keV range  [the hardness-luminosity diagram (HLD)] 
for seven sources:
{\it Z-}sources Sco~X-1, GX~340+0 (STF13)  and  $atoll$ sources GX~3+1(ST12), 4U~1820-30 [\cite{tsf13}, hereafter TSF13],  
4U~1728-34 (ST11) and BHC sources SS~433 (ST10),  4U~1630-47 (STS14). 

As it is seen  from  Fig.~\ref{HID_7object},  upper panel, these  NS and BH 
binaries trace specific tracks  in this HLD. 
We find that various types of NS LMXB subclasses can be well traced by the soft X-ray luminosity. 
More specifically, the spectral hardness HC  
of different types of NS LMXB subclasses (atolls, GX atolls, 
Sco-like {\it Z-}sources and Cyg-like {\it Z-}sources) is a function of the  luminosity $L_{3-10}$  in soft energy range from 3 to 10 keV.
 In this way a {\it Z-}source traces its {\it Z-}branches at higher mass accretion rates, while 
 {\it atolls} are observed at lower mass  accretion rate values ($\dot M$).
Note that we assume  that $L_{3-10}$ is proportional to mass accretion rate.
We should point out  that {\it atolls}   and BHs  are similar when their  luminosities are relatively low while {\it Z} sources are entirely different from BHs. 
It is interesting that BHs span almost the whole X-ray luminosity range as is covered by various typed of NS LMXB 
subclasses. Figure~\ref{HID_7object} ($bottom$ panel) shows that {\it Z} sources are more luminous than BHs in their 
soft states. 
 
It is interesting that a Cyg-like {\it Z-}source (e.g., GX~340+0) traces its 
{\it Z-}branches at some higher $\dot M$ than Sco-like {\it Z-}sources (e.g., Sco X-1). In turn, bright {\it atolls} 
(e.g., GX~3+1) trace its color-color diagrams (usually, between lower banana to upper banana) 
at some higher $\dot M$ than ordinary {\it atolls} (e.g., 4U~1728-34, 4U~1820-30). 
In this aspect it is interesting to review  and investigate the spectral properties of  XTE~J1701-462
[see Homan at el. (2010)].  This source is  a unique X-ray source
 to study  its spectral characteristics along outburst phases with different properties of NS systems versus 
its luminosity. It shows all the flavors of  NS systems: firstly appears as a {\it Z} source, then it shows features 
of  {\it atoll} source and  finally  it fades.  Homan at el. (2010) found  that an initial Cyg-like behavior followed by Sco-like as the luminosity fell. Thus they concluded  that Cyg/Sco like  types depend on luminosity. 
They also compared the observed tracks for XTE~J1701-462 with those found in the various NS LMXB subclasses and   suggested that  for the {\it Z} and {\it atolls}  can be linked through changes in a single variable parameter, namely the mass accretion rate for  the wide variety in behavior observed in NS-LXMBs with 
different luminosities.

We  also compare the electron temperature variations along spectral state changes for different NS LMXB 
subclasses (see Fig.~\ref{HID_7object}, bottom panel). 
As one can  see that $kT_e$ is  also traced by $L_{3-10}$.   The electron temperature  $kT_e$ in {\it atolls} is usually related  to relatively  lower luminosity, while that in {\it Z-}sources  corresponds  to 
 the luminosity very close the Eddington one. In fact,  this diagram demonstrates a clear difference between {\it atoll}
 and {\it Z-}sources.  The similar correlations  were found by Lin et al. (2009) and Homan et al. (2010) 
 for  the case of XTE~J1701-462. They claimed that for 2006 -- 2007 outburst events the luminosity  $L_{3-10}$
 was a good parameter to track a gradual evolution of the CCD and HID 
tracks and allowing them to relate the 
observed properties to other NSs. In particular, they detected XTE~J1701-462 with {\it Z-}source properties 
in terms of its CCD and HID 
when the  source
was at high luminosity phase.  
On the hand  XTE~J1701-462 exhibited  its {\it atoll} properties when its luminosity becomes smaller.

\section{Conclusions \label{summary}}


In this Paper, we study the correlations between spectral, timing properties, and mass accretion rate observed in
X-ray source Sco~X-1  with {\it RXTE}.
  We find that the broad-band energy 
spectra of Sco~X-1 during all 
{\it Z}-states 
can be adequately reproduced by  an additive model, consistent of 
two Comptonized 
components 
with different  {\it seed} photon temperatures (for example, $T_{s1} =0.7$ keV  and $T_{s2} =1.6$ keV).
We should also  include  an iron-line represented by a  {\it Gaussian} component to the model 
 in order to obtain  reliable fits.
 
\begin{figure}[ptbptbptb]
\includegraphics[scale=1.10,angle=0]{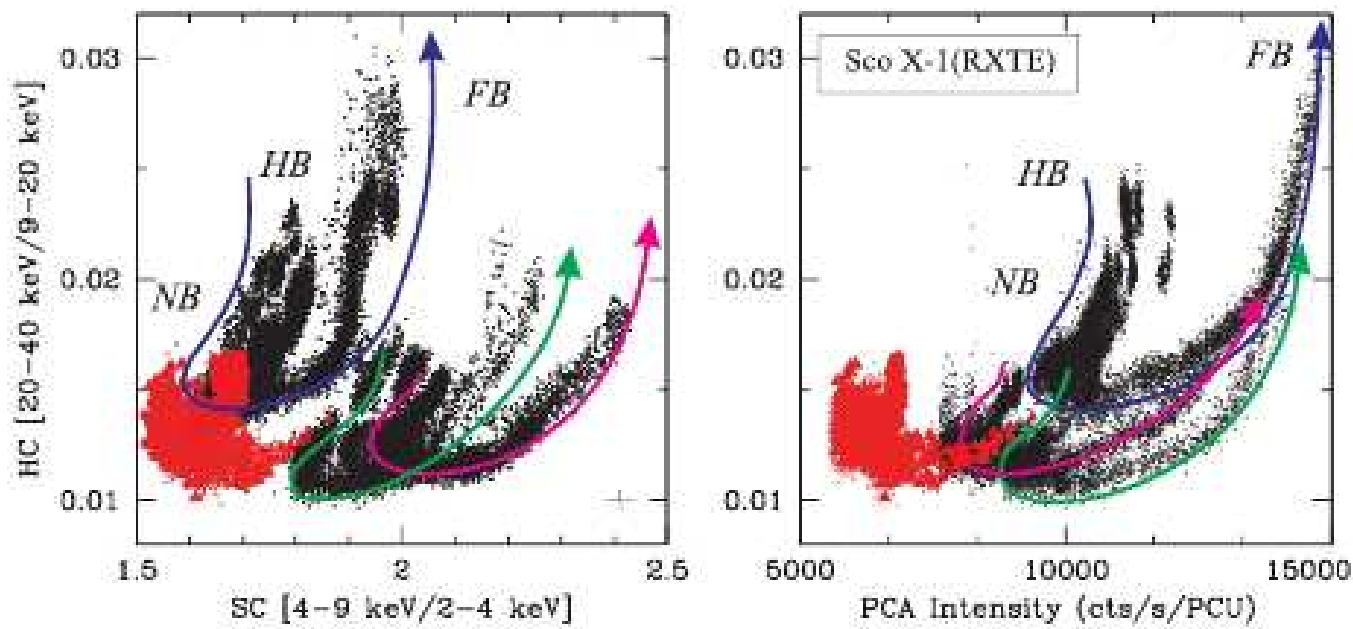}
\caption{
CCDs (left panel) and HIDs (right panel) for all observations of Sco~X-1 used in our analysis, with  
bin size 16 s. 
 The typical error bars for the colors are shown in the bottom right corner of left panel; errors
in the intensity are negligible.  
The sets with strong flaring activity are marked by black points, while the sets with 
reduced flaring activity highlighted by $red$ points. Three typical tracks with directions 
of the HB$\to$NB$\to$FB transitions for strong flaring activity sets are indicated by 
corresponding arrows. Here the spectral branches have been indicated:  the flaring branch 
(FB), normal branch (NB) and horizontal branch (HB) for $blue$ line track. Different positions 
of transitional tracks in these diagrams demonstrate a secular shifting of Sco~X-1.
}
\label{HID_Sco}
\end{figure}

\begin{figure}[ptbptbptb]
\includegraphics[scale=0.95,angle=0]{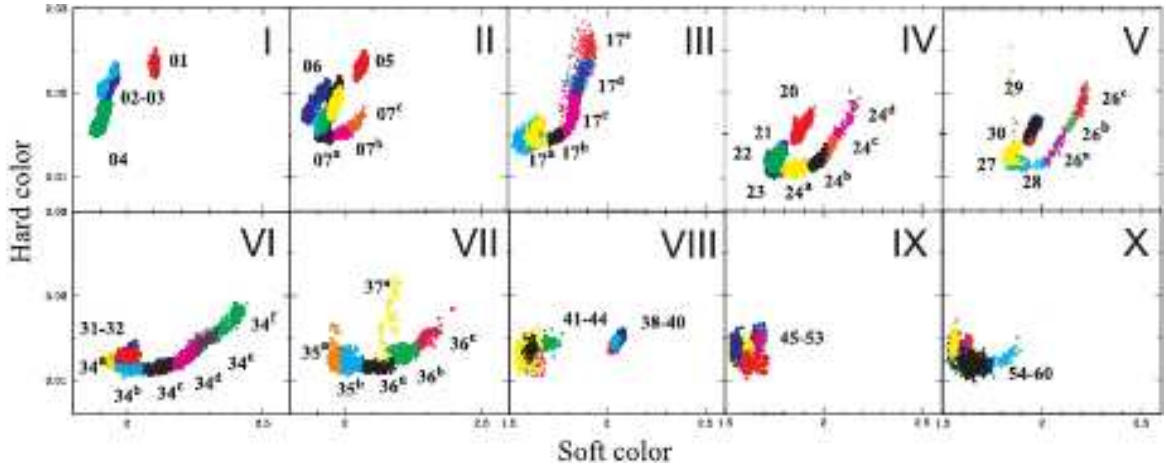}
\caption{
CCDs extracted using {\it RXTE} data sets.
Different colors point to different selected regions from which the CCD-resolved spectra
were produced. We label the spectra 
in accordance with Table 2. 
}
\label{CCD_Sco}
\end{figure}

\begin{figure}[ptbptbptb]
\includegraphics[scale=0.99,angle=0]{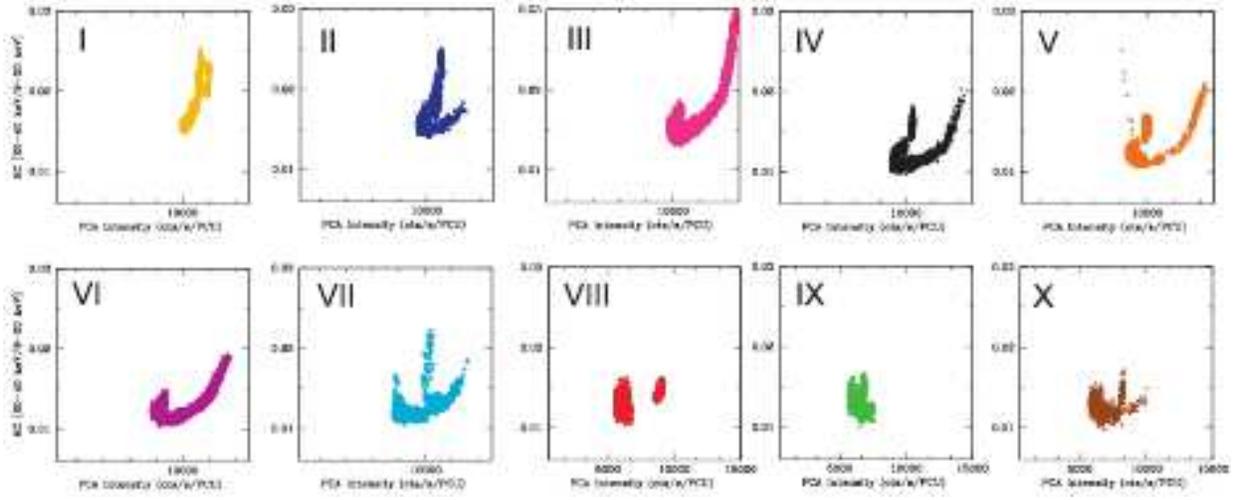}
\caption{
Evolution of the HID tracks of Sco~X-1 as a function of the count rate in [9 -- 20 keV] energy band. 
Data groups (I-X) correspond to the same data  groups as shown in Fig.~\ref{CCD_Sco}.
}
\label{all_HID_Sco}
\end{figure}

\begin{figure}[ptbptbptb]
\includegraphics[scale=0.8, angle=0]{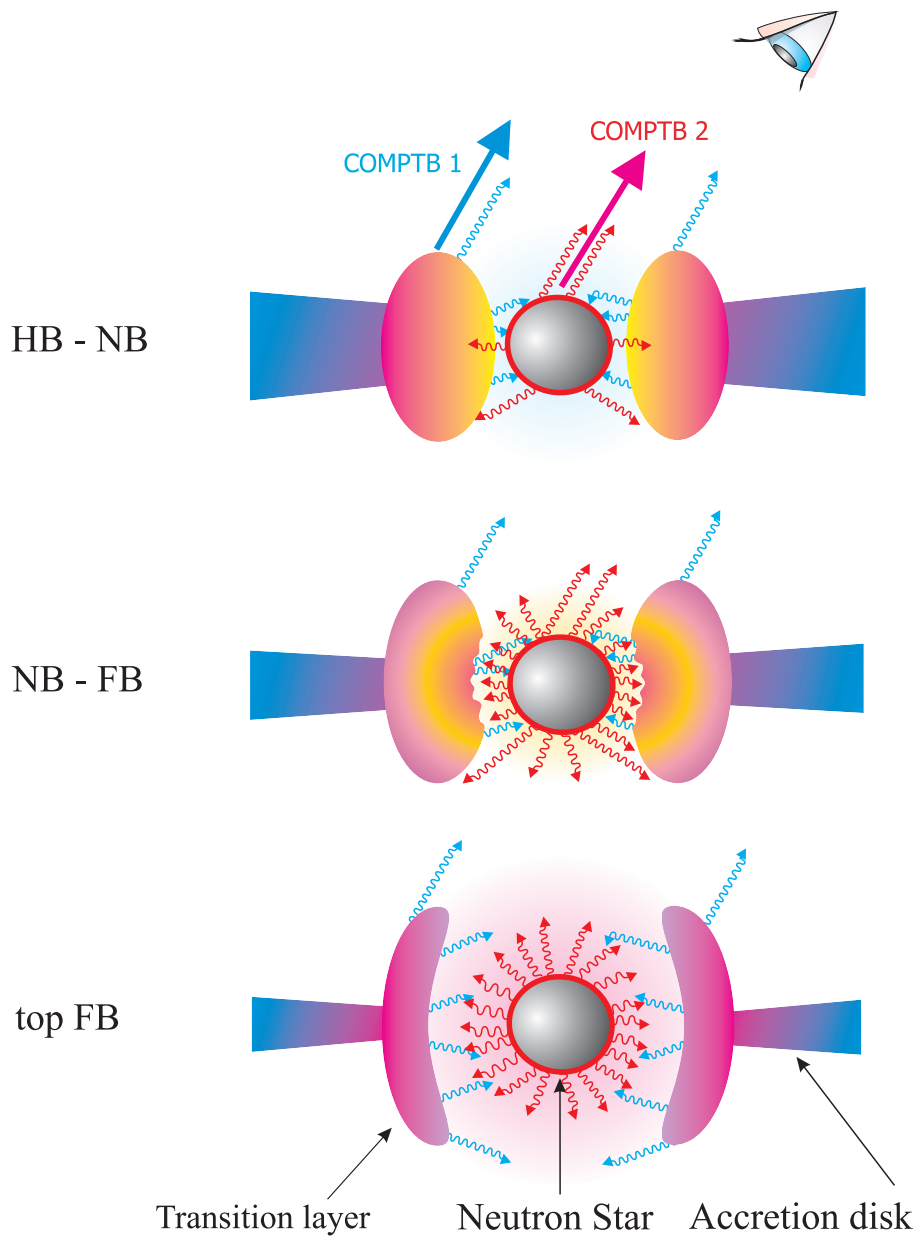}
\caption
{A suggested  geometry of Sco~X-1.   Disk and neutron star soft photons are 
upscattered off   hotter plasma of the transition layer (TL)  located 
between the accretion disk and NS surface. 
Red and blue photon trajectories correspond to soft 
and hard 
photons respectively. 
Two  Comptonized components are considered.  The first one ($Comptb1$) related to  
{\it seed} (disk) photon temperature $T_{s1}\lax 1$ keV and the TL 
electron temperature $kT^{(1)}_ e$ varies from 3 keV to 180 keV.
In the second Comptonized component ({\it Comptb2}) the temperatures   $T_{s2}\sim$1.5 keV and $kT^{(2)}_ e$ are presumably  related  to the NS surface and its inner part of the TL  respectively.
During the FB ($bottom$ panel) radiation pressure near the Eddington accretion rate disrupts 
the inner  part of the accretion disk.
}
\label{geometry}
\end{figure}

\begin{figure}[ptbptbptb]
\includegraphics[scale=0.88,angle=0]{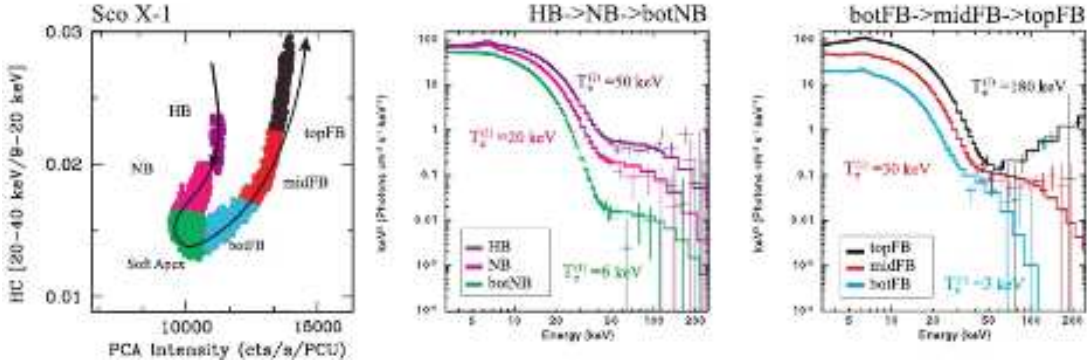}
\caption{{\it Left panel:}
Hardness-intensity diagram of Sco~X-1.
The direction of the HB$\to$NB$\to$bottomFB 
(left branch) and the bottomFB$\to$topFB  (right branch) evolution is indicated by arrows.
{\it Central panel:} 
three representative spectra
for different states along the {\it left}
{\it Z} branch [HB$\to$NB (hard apex)$\to$botNB] track. 
The spectra correspond to data 
taken from the {\it RXTE} observations 
20053-01-01-03 ({\it green}, botNB), 
20053-01-01-06$^b$ ({\it pink}, NB) 
and  20053-01-01-00 ({\it violet}, HB). 
{\it Right panel:}  Spectra   
along  
{\it Z}  track from {botFB$\to$topFB}. 
The corresponding data are taken from the {\it RXTE} observations 
20053-01-01-02$^e$ ($bright~blue$, 
{botFB}), 
20053-01-02-01$^a$ ({\it red}, midFB) 
and  20053-01-02-04$^{a-b}$ ({\it black}, topFB). 
}
\label{sp_compar_xte}
\end{figure}

\begin{figure}[ptbptbptb]
\includegraphics[scale=1.03,angle=0]{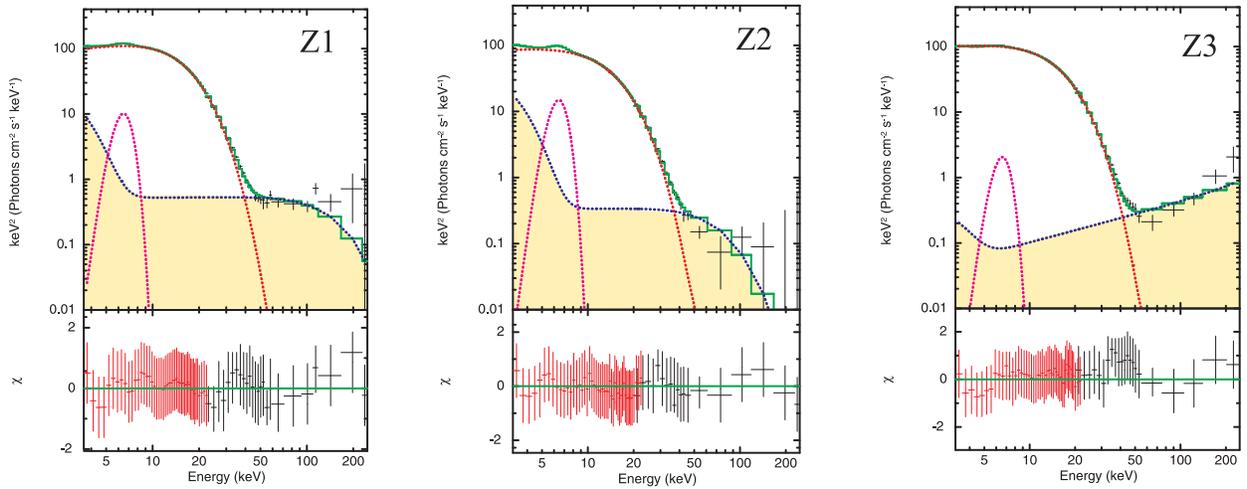}
\caption{
Three representative spectra
for different states along {\it Z-}track. 
Data are taken from the {\it RXTE} observations 
20053-01-01-00 ({HB}, left),
20053-01-01-06$^b$ ({NB}, center), 
and 20053-01-01-02$^e$ ({FB}, right).
The data are shown by black crosses and  
 the spectral model components are displayed  by dashed $blue$, $red$ and $purple$ lines for the $Comptb1$, $Comptb2$, 
 and $Gaussian$ respectively. Yellow shaded areas demonstrate an evolution of the $Comptb1$ component 
during evolution 
between the {HB}, {NB} and {FB} states. 
}
\label{Zsp_compar_RXTE}
\end{figure}

\begin{figure}[ptbptbptb]
\includegraphics[scale=0.8,angle=0]{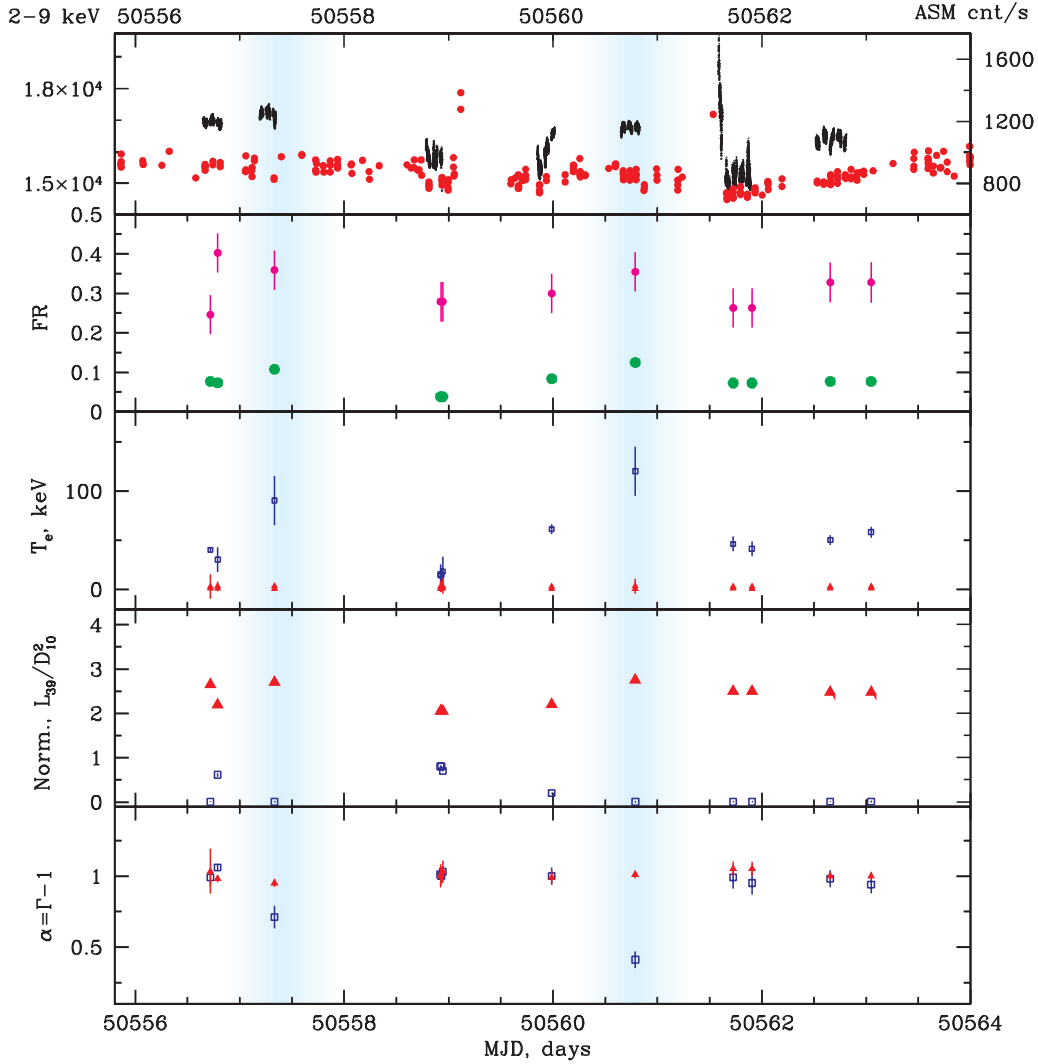}
\caption{{\it From Top to Bottom:}
Evolutions of  count rate [2 -- 9 keV] in counts s$^{-1}$ with 16~s time resolution ($black$ points, see $left$ scale  axis) and 
ASM count rate ($red$ points, see $right$ scale axis), 
the {\it flux ratio} coefficient FR [10-50 keV]/[3-10 keV], the electron temperatures $kT^{(1)}_e$ ($blue$) and 
$kT^{(2)}_e$ ($red$) 
in keV,  
$blackbody$ normalizations of the $Comptb1$ and $Comptb2$ ($blue$ and $red$  respectively)   
and  the spectral indices $\alpha_1$ and $\alpha_2$ ($\alpha=\Gamma-1$) ($blue$ and $red$) 
for the $Comptb1$ and $Comptb2$  components, respectively, during MJD 50555 -- 50565 
($R1$ set). 
The flaring phases of the light curve with the high TL electron temperature ($kT^{(2)}_e>60$ keV) and the low spectral index 
($\Gamma_1=\alpha+1<2$) 
are marked with blue vertical strips. 
}
\label{lc_r1}
\end{figure}

\begin{figure}[ptbptbptb]
\includegraphics[scale=0.9,angle=0]{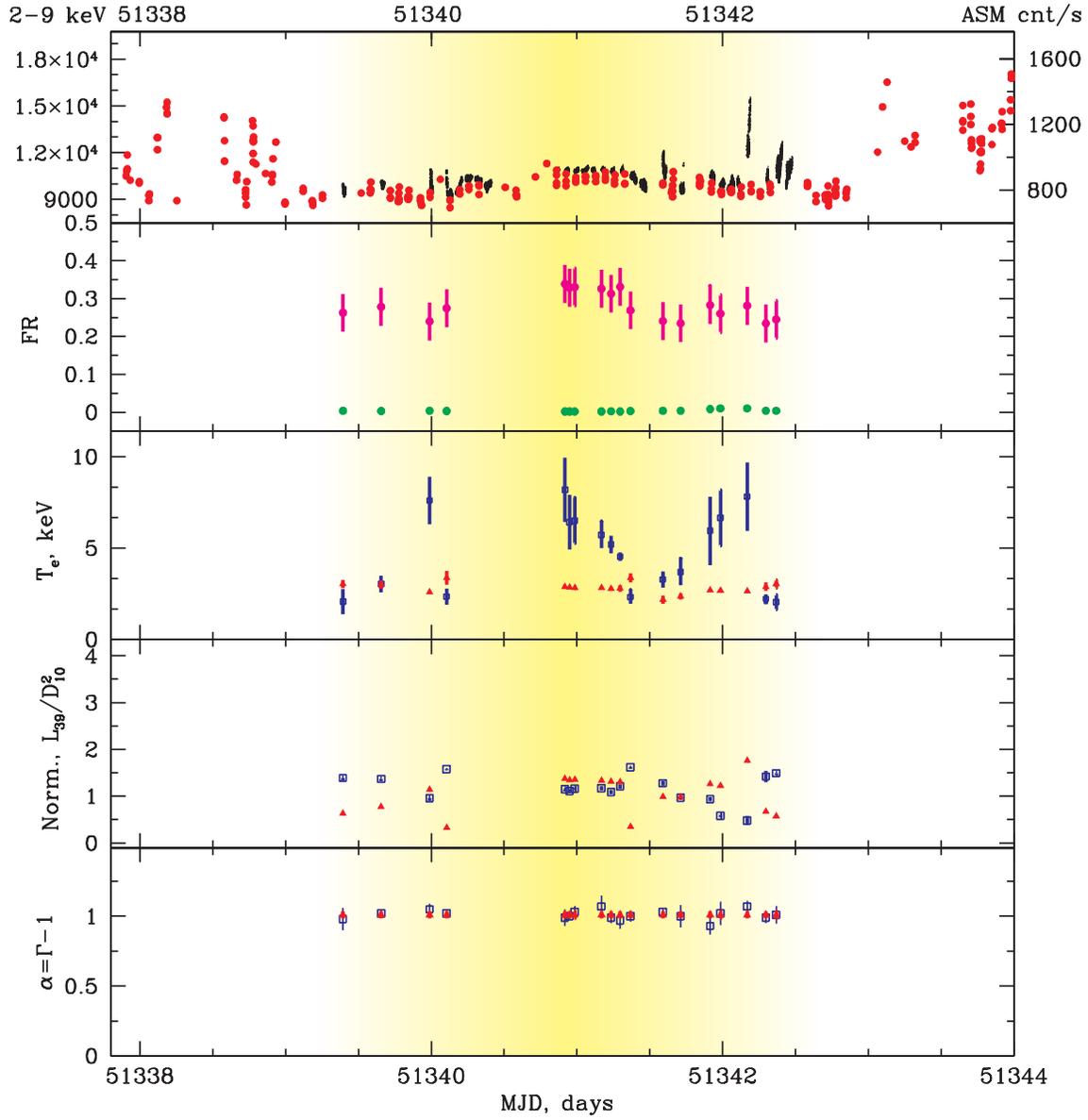} 
\caption{
Same to  that presented in Fig.~\ref{lc_r1} but  for 
MJD 51336 -- 51345 ($R5$ set) with reduced flaring activity
(marked by yellow vertical strip).
}
\label{lc_low}
\end{figure}

\begin{figure}[ptbptbptb]
\includegraphics[scale=0.9,angle=0]{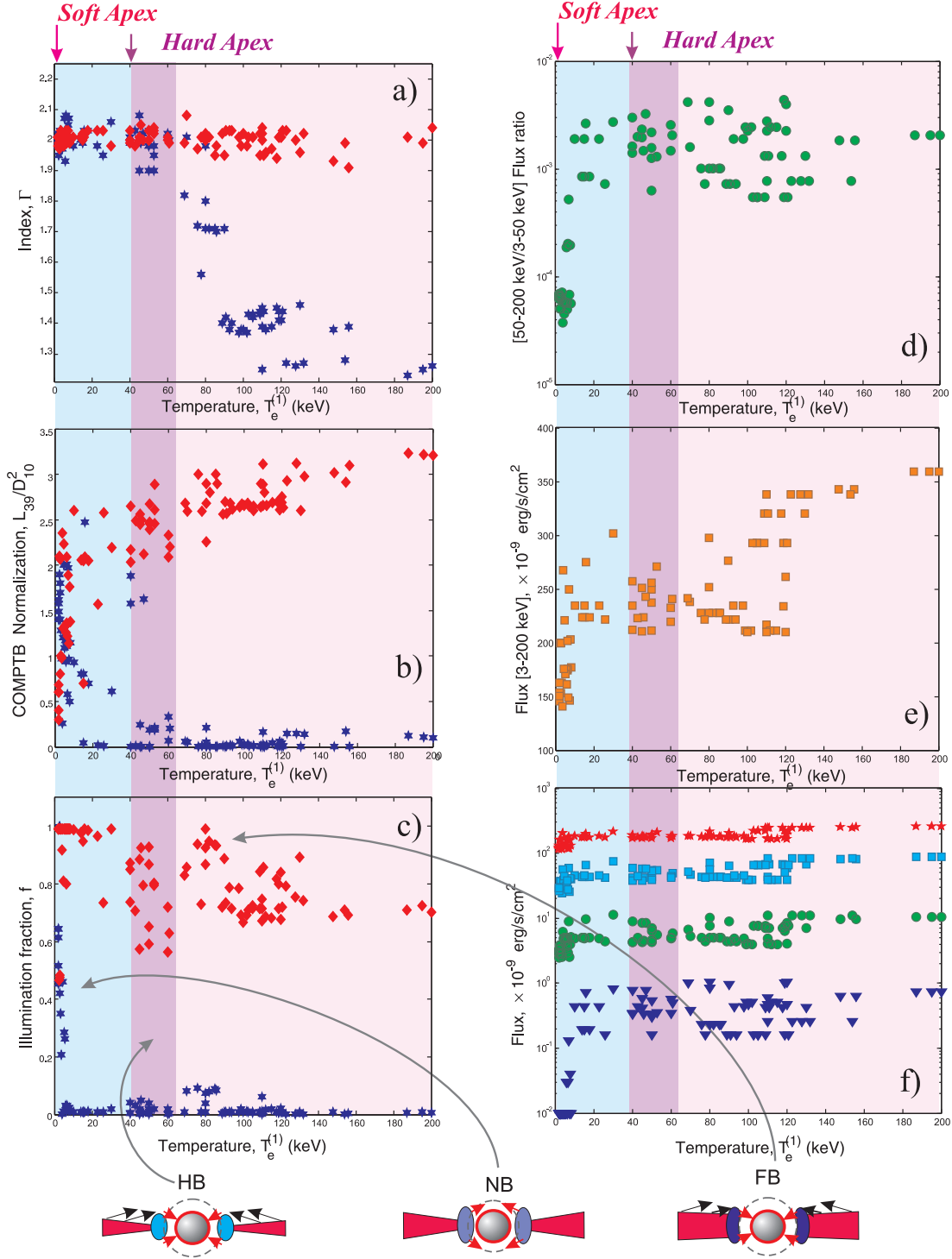}
\caption{Evolution of spectral parameters vs   $T_e^{(1)}$.  
{\it Left column}: 
The photon index (panel $a$), 
COMPTB Normalization ($b$) and 
illumination fraction $f$ ($c$) for  the {\it hard} Comptonized component ({\it Com 1}) ({\it blue} points)  and 
{\it soft}   Comptonized component ($Com 2$) ($red$ points), vs 
$T_e^{(1)}$.
{\it Right column}:  
($d$) Hardness [50 -- 250 keV]/[3 -- 50 keV]  ({\it green} points), ($e$)
the (3 -- 250 keV) flux  ({\it yellow} points) and 
the  (3 -- 10 keV), (10 -- 20 keV), (20 -- 50 keV) and (50 -- 250 keV) fluxes  (panel $f$)
vs 
$T_e^{(1)}$ (in keV) from the top to the bottom respectively.
}
\label{te1-Zstate}
\end{figure}

\begin{figure}[ptbptbptb]
\includegraphics[scale=0.7,angle=0]{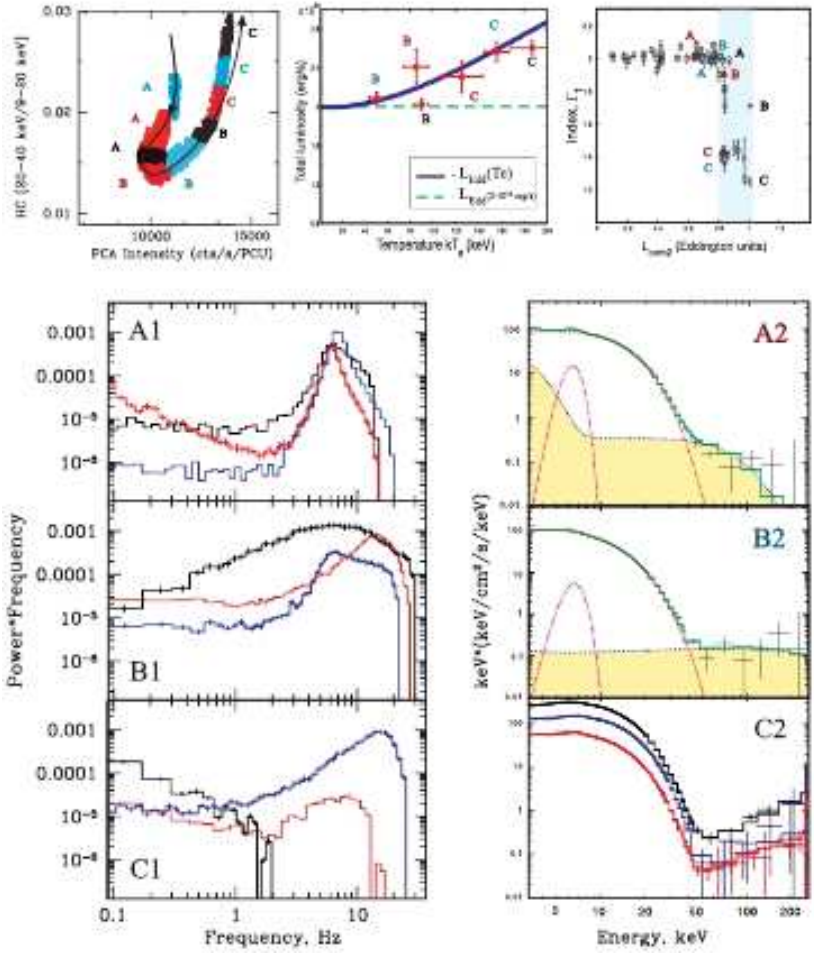}
\caption{
{\it Top}: 
HID ($left$), 
total luminosity vs.  $kT_e^{(1)}$ ($center$) and $\Gamma_1$ vs. 
 $L_{com2}$ 
in Eddington units ($right$). 
Red/blue/black points A, B, and C mark moments at 
MJD = 50557.2/50556.6/50819.6$^a$, 50818.7$^a$/50817.8/50558.7$^a$ and 50816.9$^c$/50817.8$^b$/50820.9$^e$ 
related to different 
phases of {\it Z}-track (see Table 2). 
$Bottom$: 
PDSs for 3-13 keV  band ($left$ column) are plotted along with the spectra
($right$ column) 
for A ($red$), B ($blue$) and C 
points  of X-ray flux ratio vs. flux diagram (see {\it left upper} panel). 
The data are shown by black points and  
 the spectral model components are displayed by dashed $red$, $blue$ and $purple$ lines for the  $Comptb1$, $Comptb2$ 
and $Gaussian$ respectively.  Yellow shaded areas demonstrate  an evolution of
the $Comptb1$ component during evolution between the HB and FB.  The normalization factors of 2 and 0.5 were  applied 
for 20053-01-02-01$^b$ ($blue$) and 20053-01-02-00$^c$ ($red$) spectra respectively (see panel C2).
}
\label{PDS}
\end{figure}

\begin{figure}[ptbptbptb]                        	
\includegraphics[scale=1.10, angle=0]{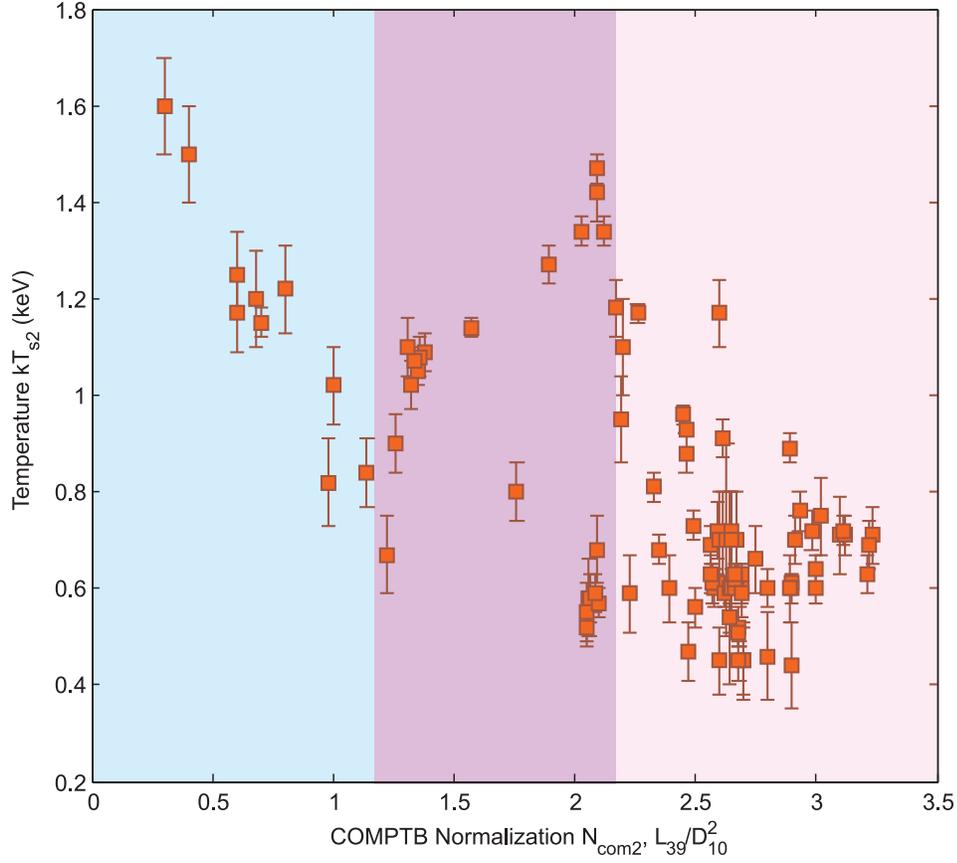}
\caption{
The seed photon temperature $kT_{s2}$ (in keV), 
using  our spectral model $wabs*(Comptb1+Comptb2+Gaussian)$,   
  is plotted as a function of 
the COMPTB Normalization of the $soft$ Comptonized component  $N_{com2}$ (in $L_{39}/D^2_{10}$ units).
The colors of vertical strips correspond to that of Fig.~\ref{te1-Zstate}.
}
\label{norm2-Zstate}
\end{figure}

\begin{figure}[ptbptbptb]
\includegraphics[scale=1.1,angle=0]{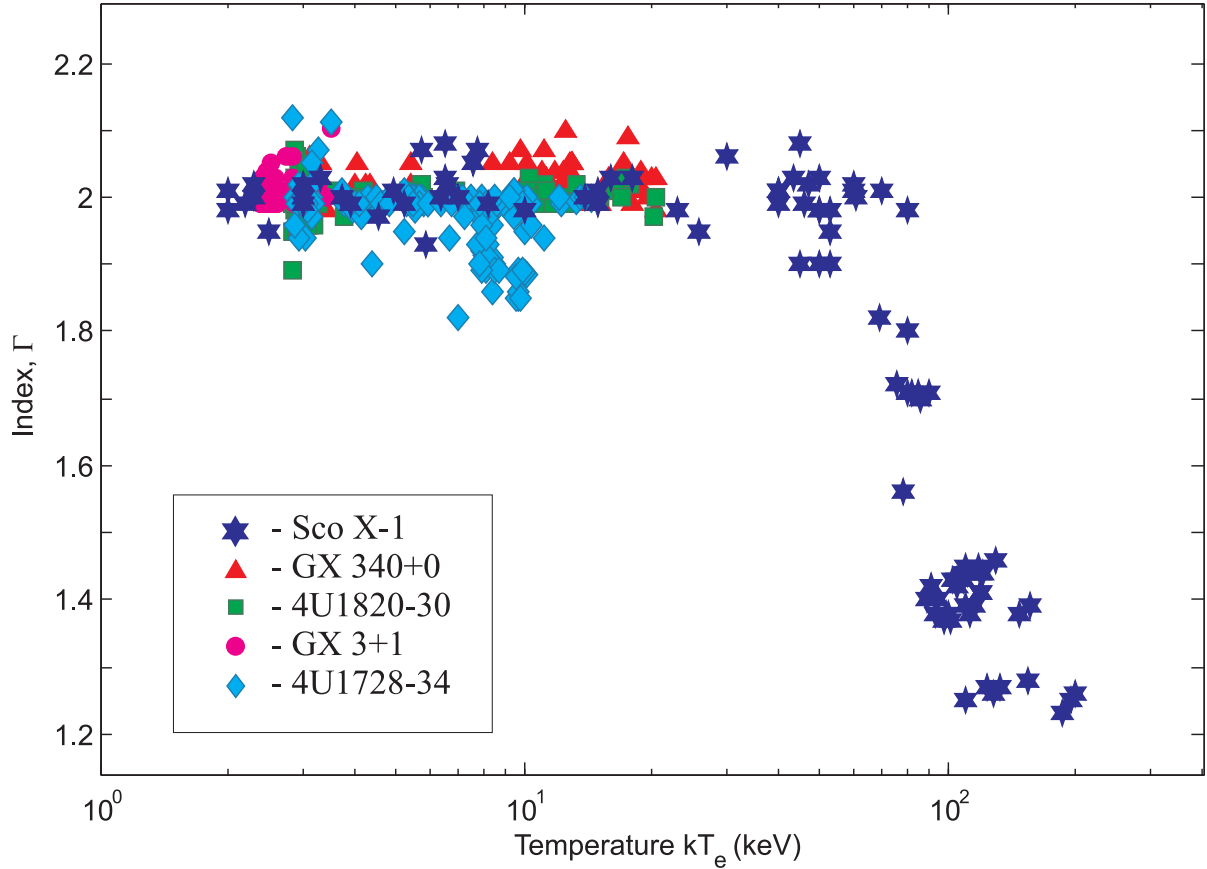} 
\caption{
The photon 
index $\Gamma$ 
vs
$kT_e$ 
for {\it Z}-sources Sco~X-1 ($blue$ stars), GX~340+0 ($red$ triangles, taken from STF13) and $atoll$ sources 4U~1728-34 
($bright~blue$ diamonds, taken from ST11), GX~3+1 ($pink$ circles, taken from ST12) and 4U~1820-30 ($green$ squares, 
taken from TSF13). 
}
\label{gam_te_5obj}
\end{figure}

\begin{figure}[ptbptbptb]
\includegraphics[scale=0.75,angle=0]{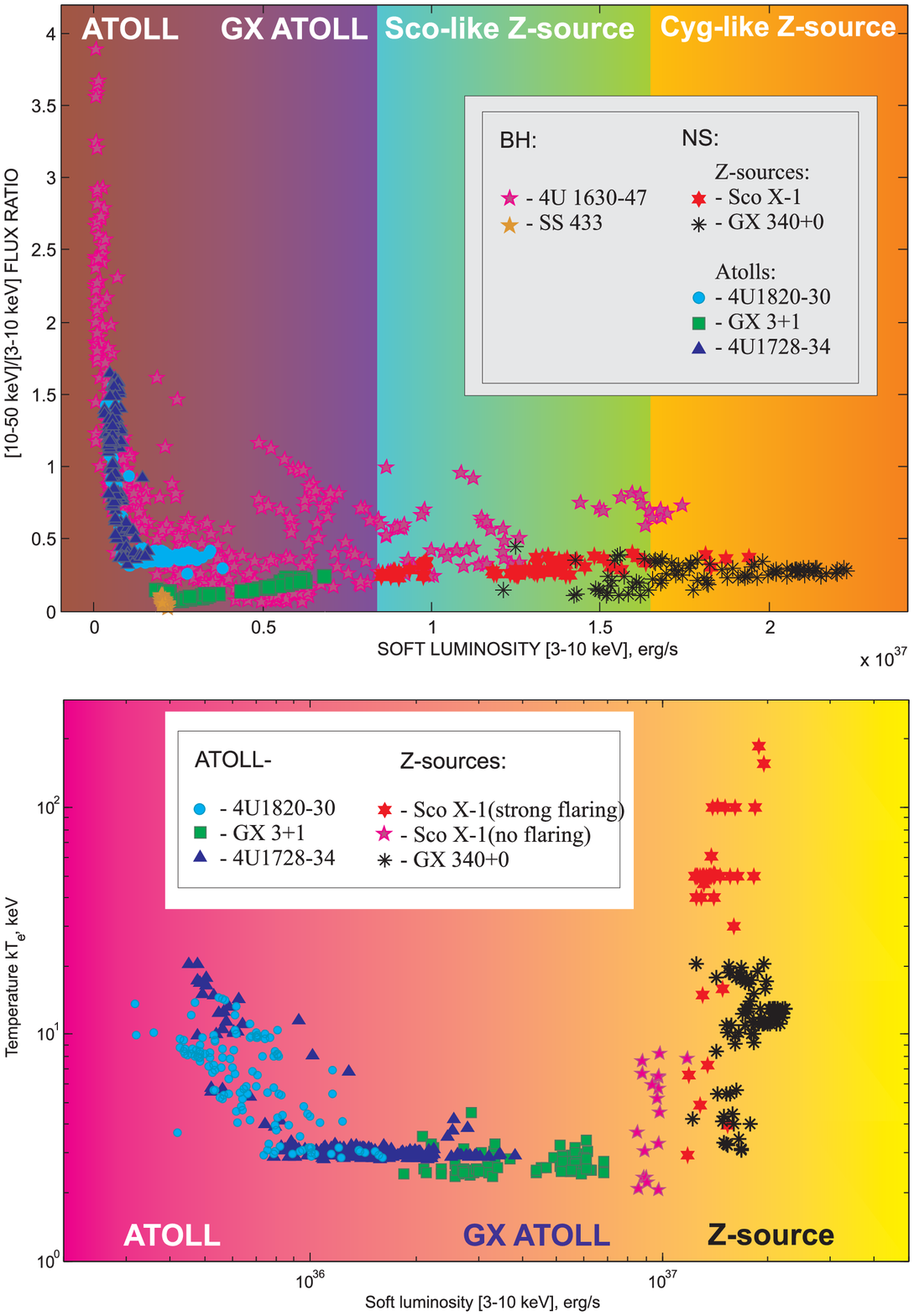}
\caption{
{\it Upper panel:} Spectral hardness (10-50 keV/3-50 keV) vs luminosity 
in 3-10 keV range, 
for the  {\it Z} sources Sco X-1 (red), 
GX~340+0 (black, from STF13), {\it atolls} 4U 1728-34 (blue,  from ST11), GX 3+1 (green, from ST12),  
4U~1820-30 (green, from TSF13) and BHCs 4U~1630-47 (pink,
 from STS14), SS~433 (hazel, from ST10).  
{\it Bottom panel:} The electron temperature $kT_e$ vs luminosity in 3-10 keV range, 
for the  {\it Z} sources Sco~X-1 (red), GX~340+0 (black,  from STF13), {\it atolls} 4U~1728-34 (blue,  from ST11), GX~3+1 
(green,  from ST12) and  4U~1820-30 (bright blue,  from TSF13). 
}
\label{HID_7object}
\end{figure}

 Our approach to modeling the X-ray spectra of Sco~X-1 allows us to separate the contributions of  two distinct   zones of the spectral formation  related to 
the $hard$ and $soft$ Comptonized components 
along the {\it  Z} track which is possibly driven by mass accretion rate $\dot M$.  In fact,  
the observed soft photon luminosity which is  determined by
{\it normalization} parameters of the  Comptonized components (see Eq. \ref{comptb_norm}) is proportional to $\dot M$.  
We observe an increase of the $N_{Com1}$ and $N_{Com2}$ (Fig.~\ref{te1-Zstate}) 
from  the $soft~apex$ to the $hard~apex$  in the CCD 
(see Fig. \ref{HID_Sco})  
which can indicate to an increase of mass accretion rate $\dot M$ from the {\it lower} normal branch  to 
the {\it upper} normal branch.
 The normalization parameter $N_{Com2}$ is related to the soft (NS)  Comptonized component,  
and is related  a total mass accretion rate $\dot M_{tot}$.  One can see from Fig. \ref{te1-Zstate}, panel b there,  how the total mass accretion rate, or  $N_{Com2}$, increases when  $kT_e$ increases and using Fig. \ref{HID_Sco} one can also see how the hardness ratio increases from the  {\it lower} FB  to the {\it upper} FB.
We should also point out that a comparative analysis using our model shows that various NS LMXBs are also  well traced by soft   luminosity in the energy range from 3 to 10 keV.

Our spectral analysis 
also allows us to reveal
the stability of both photon indices $\Gamma_1$ and $\Gamma_2$ 
around 2 during the HB/NB/botFB states, while the decrease of $\Gamma_1$ is observed  in the mid-topFB. 
during this mid-topFB segment.  
We interpret the detected quasi-stability of the indices of Comptonized components 
about a value of 2 during the HB -- NB -- botFB 
in the framework of the model in which the gravitational energy release  occurs in the  TL and when that is much higher  than the soft (disk) flux.
 This index stability phase 
is now established for the Comptonized spectral components of Sco~X-1 over 
the HB -- NB -- botFB. This result is similar to those were previously found in  the  $atoll$ sources 
4U~1728-34, GX~3+1, 4U~1820-30 and  {\it Z-}source GX~340+0 through all spectral states. {\it The new index reduction   phase detected over the FB in Sco~X-1 for the hard $Comptonized$ component} we interpret in the framework of the  model in which the gravitational energy release takes place only in some outer part of the transition layer.  But in its lower part the soft photon illumination increases (mainly by NS soft photons)  which is followed by a decrease of the plasma temperature. As a result   the electron cross-section increases and   the radiation pressure reaches  its critical value  and thus it stops the accretion flow   
(see \S \ref{index reduction}).  


Note during this index reduction stage the  X-ray spectrum of Sco~X-1 exhibits  
    the  electron temperature $T_e^{(1)}$ increase  from 60 keV to 180 keV. 
 It is important to emphasize that in BH sources the index increases up to a saturation level  when the luminosity increases. This behavior is opposite to that seen in Sco X-1.

We are very grateful to the referee whose constructive suggestions help us to improve the paper quality.

\newpage

%
%
\newpage
\begin{deluxetable}{llll}
\tablewidth{0in}
\tabletypesize{\scriptsize}
    \tablecaption{The list of groups of {\it RXTE} observation of Sco~X-1}
    \renewcommand{\arraystretch}{1.2}
\tablehead{Number of set  & Dates, MJD & RXTE Proposal ID&  Dates UT  \\
                          &            &                 &                }
 \startdata
R1  &    50227-50228; 50522, 50815  & 10061$^1$ & 1996 May 24 -- 25, 1997 March 15; 1998 Jan 2            \\
R2  &    50556-50562; 50816-50821   & 20053$^{1,2,3}$ & 1997 April 18 -- 24; 1998 January 3 -- 8                \\
R3  &    50820-50821; 50963-50999   & 30036$^{1,2,3,4}$; 30035$^{1,2}$ & 1998 Jan 7 -- 8; May  30 -- July 5 \\ 
R4  &    51186-51194                & 40020$^2$ & 1999 Jan  1 -- 16                                       \\
R5  &    51339-51342                & 40706$^{1,3,5}$ & 1999 Jun  10 -- 13$^{\dagger}$           \\
R6  &    52413.9-52414              & 70015$^2$ & 2002 May 19 21:55:44 -- 23:45:28$^{\dagger}$                        \\
      \hline
      \enddata
    \label{tab:par_bbody}
References:
(1) \cite{D'Amico01};  
(2) \cite{D'Ai07}; 
(3) \cite{Church06}, \cite{Church12};  
(4) \cite{Barnard03}; 
(5) \cite{Bradshaw03};  
{$^\dagger$} reduced flaring sets
\end{deluxetable}

%
%

%
%
\newpage
\bigskip
\begin{deluxetable}{llccccccccccccc}
\rotate
\tablewidth{0in}
\tabletypesize{\scriptsize}
    \tablecaption{Best-fit parameters of spectral analysis of PCA+HEXTE/{\it RXTE} 
observations of Sco~X-1 in 3 -- 250~keV energy range$^{\dagger}$. Adopted model: {\it wabs*(Comptb1 + Comptb2 + Gaussian)}.  
Parameter errors are given at 90\% confidence level.}
    \renewcommand{\arraystretch}{1.2}
 \tablehead                                                                                                          
{No. & Observational & MJD, & $\alpha_1=$  & $kT^{(1)}_e,$ & $\log(A_1)$& $N_{Com1}^{\dagger\dagger\dagger}$ & $kT_{s2}$, & $\alpha_2=$  & $kT^{(2)}_e,$ & $log(A_2)$ & N$_{Com2}^{\dagger\dagger\dagger}$ &E$_{line}$,& $N_{line}^{\dagger\dagger\dagger}$ &  $\chi^2_{red}$ (d.o.f.) \\
     & ID            & day  & $\Gamma_1-1$ & keV        &           &                                    &       keV                              & $\Gamma_2-1$ & keV        &                                    & keV       &                                    &                         &                                     }
 \startdata
01 & 10061-01-01-00  & 50227.865 & 1.03(2) & 16(4)  & -2.0(5)  & 2.47(3) & 1.42(6) & 1.01(3) & 2.94(1) & 2.00$^{\dagger\dagger}$  & 2.09(1) & 5.99(4) & 1.18(5) & 1.21(87) \\
02 & 10061-01-01-01  & 50228.003 & 1.00(3) &  7(2)  & -1.6(2)  & 1.97(8) & 1.27(4) & 1.00(1) & 2.97(4) & 2.00$^{\dagger\dagger}$  & 1.89(2) & 6.02(6) & 0.76(7) & 0.93(87) \\
03 & 10061-01-02-00  & 50522.766 & 0.98(4) & 80(3)  & -1.40(1) & 0.21(7) & 1.17(2) & 1.00(2) & 3.05(1) & 2.00$^{\dagger\dagger}$  & 2.26(1) & 6.12(6) & 0.80(9) & 1.14(87) \\
04 & 20053-01-02-05  & 50556.570 & 0.99(3) & 40(1)  & -1.36(7) & 0.004(2)& 0.72(8) & 1.03(6) & 2.89(5) & 0.84(5) & 2.65(9) & 6.79(7) & 0.22(7) & 1.02(85) \\
05 & 20053-01-01-00  & 50556.639 & 1.06(1) & 30(5)  & -1.69(8) & 0.61(8) & 0.95(9) & 0.98(1) & 2.99(2) & 2.00$^{\dagger\dagger}$ & 2.19(4) & 6.3(2)  & 0.51(9) & 1.16(87) \\
06 & 20053-01-01-01  & 50557.183 & 0.71(4) & 90(10) & -2.01(4) & 0.004(1)& 0.45(7) & 0.95(1) & 3.01(1) & 0.9(1)  & 2.7(1)  & 6.46(9) & 0.30(5) & 1.32(87) \\
07$^a$ & 20053-01-01-02$^a$  & 50558.774 & 1.01(2) & 15(4)  & -2.0(1)  & 0.8(2)  & 0.55(6) & 1.00(3) & 2.79(2) & 1.27(5) & 2.05(4) & 6.39(8) & 0.72(7) & 1.02(87) \\
07$^b$ & 20053-01-01-02$^b$  & ..... & 1.00(3) & 14(3)  & -2.1(2)  & 0.8(1)  & 0.58(8) & 1.01(2) & 2.82(2) & 1.63(8) & 2.06(7) & 6.54(7) & 0.83(7) & 1.04(87) \\
07$^c$ & 20053-01-01-02$^c$  & ..... & 1.03(2) & 18(6)  & -1.9(1)  & 0.7(2)  & 0.52(4) & 1.03(3) & 2.93(3) & 1.82(9) & 2.05(6) & 6.29(9) & 0.91(7) & 1.01(87) \\
08 & 20053-01-01-03  & 50559.841 & 1.00(3) & 61(2)  & -2.34(8) & 0.2(1)  & 1.1(1)  & 0.99(2) & 2.86(1) & 0.23(8) & 2.2(1)  & 6.4(1)  & 0.59(6) & 1.19(87) \\
09 & 20053-01-01-04  & 50560.641 & 0.41(3) & 120(10)& -1.68(7) & 0.001(1)& 0.66(7) & 1.01(1) & 3.03(3) & 0.72(7) & 2.75(2) & 6.46(9) & 0.10(2) & 1.03(87) \\
10 & 20053-01-01-05  & 50561.576 & 0.99(4) & 46(3)  & -1.29(3) & 0.003(1)& 0.56(4) & 1.05(2) & 2.86(1) & 1.10(5) & 2.50(3) & 6.70(5) & 0.44(5) & 1.16(85) \\
11 & 20053-01-01-06  & 50562.509 & 0.98(3) & 50(2)  & -1.90(4) & 0.007(2)& 0.47(6) & 1.00(1) & 2.93(1) & 1.11(6) & 2.47(2) & 6.93(9) & 0.37(6) & 0.97(85) \\
12 & 10061-01-03-00  & 50815.621 & 1.01(2) & 70(5)  & -1.05(8) & 0.05(1) & 0.72(6) & 1.08(3) & 2.92(3) & 0.78(6) & 2.59(4) & 6.78(8) & 0.33(7) & 1.23(87) \\
13$^a$ & 20053-01-02-00$^a$  & 50816.887 & 0.44(6) & 120(10)& -2.0(9)  & 0.002(1)& 0.44(9) & 0.95(1) & 2.76(1) & 0.73(4) & 2.9(1)  & 6.72(7) & 0.75(6) & 1.25(85) \\
13$^b$ & 20053-01-02-00$^b$  & ..... & 0.45(5) & 120(10)& -2.0(9)  & 0.001(2)& 0.45(8) & 0.94(2) & 2.77(1) & 0.65(3) & 2.7(2)  & 6.67(4) & 0.73(4) & 1.26(85) \\
13$^c$ & 20053-01-02-00$^c$  & ..... & 0.44(4) & 110(10)& -2.0(9)  & 0.001(1)& 0.46(9) & 0.95(1) & 2.65(2) & 0.78(4) & 2.8(1)  & 6.52(8) & 0.74(6) & 1.23(85) \\
13$^d$ & 20053-01-02-00$^d$  & ..... & 0.46(6) & 120(10)& -2.0(9)  & 0.001(1)& 0.45(7) & 0.96(1) & 2.93(1) & 0.92(1) & 2.6(3)  & 6.87(9) & 0.72(3) & 1.26(85) \\
14$^a$ & 20053-01-02-01$^a$  & 50817.931 & 0.39(7) & 156(8) & -2.3(4)  & 0.19(9) & 0.71(8) & 0.91(2) & 2.85(6) & 0.35(7) & 3.10(2) & 6.96(8) & 0.45(5) & 1.16(85) \\
14$^b$ & 20053-01-02-01$^b$  & ..... & 0.39(7) & 149(7) & -2.0(4)  & 0.17(8) & 0.75(8) & 0.93(2) & 2.93(5) & 0.34(6) & 3.02(4) & 7.10(9) & 0.44(6) & 1.14(85) \\
15 & 20053-01-02-020 & 50818.773 & 0.90(4) & 50(5)  & -1.39(6) & 0.003(1)& 0.7(1)  & 1.00(4) & 2.93(4) & 0.82(5) & 2.67(8) & 6.90(8) & 0.2(1)  & 1.00(85) \\
16$^a$ & 20053-01-02-030$^a$ & 50819.623 & 0.71(2) & 85(5)  & -1.01(5) & 0.001(1)& 0.60(3) & 0.95(2) & 2.85(1) & -0.17(5)& 3.0(2)  & 7.02(9) & 0.13(3) & 1.06(85) \\
16$^b$ & 20053-01-02-030$^b$ & ..... & 0.72(1) & 76(4)  & -1.00(6) & 0.001(1)& 0.64(3) & 0.97(1) & 2.81(2) & -0.18(7)& 3.0(1)  & 6.93(4) & 0.14(2) & 0.08(85) \\
16$^c$ & 20053-01-02-030$^c$ & ..... & 0.71(2) & 82(9)  & -1.06(5) & 0.002(1)& 0.60(4) & 1.01(3) & 3.17(1) & -0.26(5)& 2.8(2)  & 7.01(7) & 0.15(2) & 1.01(85) \\
16$^d$ & 20053-01-02-030$^d$ & ..... & 0.70(1) & 86(5)  & -1.03(4) & 0.001(1)& 0.61(2) & 0.98(2) & 2.80(1) & -0.15(6)& 2.9(3)  & 7.02(9) & 0.14(3) & 1.06(85) \\
16$^e$ & 20053-01-02-030$^e$ & ..... & 0.71(2) & 80(7)  & -1.10(5) & 0.001(1)& 0.60(3) & 1.00(1) & 2.87(3) & -0.09(7)& 2.9(2)  & 6.95(8) & 0.14(2) & 1.11(85) \\
17$^a$ & 30036-01-01-000$^a$ & 50820.562 & 0.26(2) & 120(10)& -4.05(4) & 0.15(9) & 0.71(4) & 1.00(1) & 3.04(2) & 0.49(2) & 3.12(2) & 6.29(6) & 0.84(7) & 1.05(87) \\
17$^b$ & 30036-01-01-000$^b$ & ..... & 0.26(3) & 120(10)& -4.03(5) & 0.16(8) & 0.72(3) & 1.04(1) & 3.15(3) & 0.47(3) & 3.11(3) & 6.35(7) & 0.74(8) & 1.16(87) \\
17$^c$ & 30036-01-01-000$^c$ & ..... & 0.27(3) & 110(10)& -4.16(4) & 0.15(7) & 0.76(4) & 1.03(1) & 3.07(2) & 0.58(2) & 2.93(3) & 6.76(5) & 0.54(9) & 1.12(87) \\
17$^d$ & 30036-01-01-000$^d$ & ..... & 0.28(2) & 100(10)& -4.07(3) & 0.17(9) & 0.70(5) & 0.99(2) & 3.06(1) & 0.39(4) & 2.91(4) & 6.21(8) & 0.86(8) & 0.99(87) \\
17$^e$ & 30036-01-01-000$^e$ & ..... & 0.27(3) & 110(9) & -4.03(4) & 0.14(8) & 0.72(4) & 1.02(1) & 3.02(2) & 0.46(3) & 2.98(3) & 6.19(7) & 0.93(9) & 1.08(87) \\
18$^a$ & 20053-01-02-04$^a$  & 50820.955 & 0.23(3) & 190(10) & -2.2(3) & 0.12(8) & 0.71(6) & 1.01(2) & 2.86(1) & 0.39(5) & 3.23(8) & 7.00(9) & 0.39(7) & 1.18(85) \\ 
18$^b$ & 20053-01-02-04$^b$  & .....     & 0.25(2) & 195(9)  & -2.1(1) & 0.11(9) & 0.69(5) & 0.99(1) & 2.81(2) & 0.42(4) & 3.22(7) & 6.95(8) & 0.40(6) & 1.17(85) \\ 
18$^c$ & 20053-01-02-04$^c$  & .....     & 0.26(3) & 200(10) & -2.2(2) & 0.10(7) & 0.63(4) & 1.04(1) & 2.87(1) & 0.37(5) & 3.21(9) & 6.99(9) & 0.41(9) & 1.09(85) \\ 
19 & 30036-01-02-000 & 50821.555 & 1.01(2) & 60(10) & -2.78(5) & 0.07(2) & 1.47(3) & 1.00(1) & 2.94(2) & 0.11(4) & 2.09(2) & 6.00(5) & 0.91(8) & 1.17(87) \\
20 & 30035-01-01-00  & 50963.019 & 1.01(3) & 40(2)  & -2.25(6) & 1.58(9) & 1.18(6) & 0.99(2) & 3.08(1) & 0.75(6) & 2.17(4) & 6.4(1)  & 0.35(2) & 1.26(87) \\
21 & 30035-01-02-000 & 50964.018 & 1.08(2) & 45(5)  & -1.69(3) & 0.24(1) & 0.96(2) & 1.00(1) & 2.81(1) & 0.13(2) & 2.45(2) & 6.7(1)  & 0.30(1) & 1.19(87) \\
22 & 30035-01-05-00  & 50965.018 & 1.02(3) & 60(5)  & -1.7(1)  & 0.33(4) & 0.81(3) & 1.01(2) & 2.81(2) & 0.41(5) & 2.33(4) & 6.7(1)  & 0.30(2) & 1.15(87) \\
23 & 30035-01-03-00  & 50965.085 & 1.03(4) & 50(2)  & -1.7(2)  & 0.19(1) & 0.93(1) & 1.03(1) & 2.81(1) & 0.16(2) & 2.46(2) & 6.7(1)  & 0.31(3) & 1.25(87) \\
24$^a$ & 30035-01-06-00$^a$  & 50966.019 & 0.98(3) & 53(5)  & -1.7(4)  & 0.20(4) & 0.88(4) & 1.02(3) & 2.73(9) & 0.60(8) & 2.46(5) & 6.71(4) & 0.30(2) & 1.04(87) \\
24$^b$ & 30035-01-06-00$^b$  & ..... & 0.90(2) & 53(5)  & -1.7(4)  & 0.21(3) & 0.89(3) & 1.04(2) & 2.75(8) & 0.59(6) & 2.89(6) & 6.68(3) & 0.35(3) & 0.98(87) \\
24$^c$ & 30035-01-06-00$^c$  & ..... & 0.95(3) & 53(5)  & -1.7(4)  & 0.19(4) & 0.91(4) & 1.03(3) & 2.76(9) & 0.61(8) & 2.61(5) & 6.69(4) & 0.32(2) & 1.01(87) \\
25 & 30035-01-04-00  & 50966.083 & 0.99(4) & 3.9(1) & -3.0(9)  & 0.26(3) & 0.68(3) & 1.00(1) & 2.67(5) & 1.05(5) & 2.40(9) & 6.73(8) & 0.39(8) & 1.17(87) \\
26$^a$ & 30035-01-07-00$^a$  & 50996.268 & 1.00(1) & 3(1)   &  3.0(8)  & 1.7(2)  & 0.58(5) & 0.98(2) & 2.7(2)  & -0.03(1)& 2.63(8) & 6.71(4) & 0.36(7) & 1.31(87) \\
26$^b$ & 30035-01-07-00$^b$  & ..... & 0.95(2) & 2.5(2) &  3.0(8)  & 1.9(1)  & 0.57(3) & 0.99(1) & 2.8(1)  & -0.06(2)& 2.97(8) & 6.75(4) & 0.38(3) & 1.30(87) \\
26$^c$ & 30035-01-07-00$^c$  & ..... & 0.99(2) & 3(1)   &  3.0(8)  & 1.8(2)  & 0.59(4) & 0.97(2) & 2.7(2)  & -0.04(1)& 2.83(8) & 6.69(4) & 0.37(5) & 1.28(87) \\
27 & 30035-01-08-00  & 50997.268 & 1.01(3) & 40(5)  & -2.42(2) & 1.88(6) & 1.34(3) & 1.00(1) & 3.01(2) & 0.45(6) & 2.03(2) & 6.10(8) & 0.46(2) & 1.23(87) \\
28 & 30035-01-09-00  & 50998.268 & 0.90(4) & 45(3)  & -3.0(8)  & 0.004(1)& 0.69(4) & 1.01(1) & 3.08(1) & 0.89(8) & 2.56(6) & 6.72(4) & 0.29(1) & 1.05(87) \\
29 & 30035-01-10-00  & 50999.268 & 0.98(3) & 50(6)  & -3.0(9)  & 0.001(1)& 0.60(7) & 1.03(2) & 2.89(7) & 0.27(6) & 2.39(2) & 6.72(4) & 0.73(2) & 1.19(87) \\
30 & 30035-01-11-00  & 50997.402 & 1.02(2) & 47(8)  & -2.26(2) & 1.63(6) & 1.34(3) & 1.00(1) & 3.09(1) & 0.59(8) & 2.12(3) & 6.2(1)  & 0.34(5) & 1.15(87) \\
31 & 40020-01-01-00  & 51186.402 & 0.45(4) & 110(10)& -1.19(7) & 0.005(1)& 0.60(7) & 1.02(3) & 2.83(1) & 0.4(2)  & 2.89(9) & 6.4(1)  & 0.65(7) & 1.23(87) \\
32 & 40020-01-01-01  & 51186.920 & 1.03(2) & 43(6)  & -1.5(1)  & 0.01(1) & 0.73(3) & 0.98(2) & 2.82(2) & 0.38(4) & 2.49(3) & 7.24(6) & 0.34(5) & 1.28(87) \\
33$^a$ & 40020-01-01-020$^a$ & 51187.469 & 0.95(3) & 26(3)  & -2.0(2)  & 0.011(3)& 0.60(4) & 1.03(1) & 2.78(3) & 0.44(5) & 2.58(4) & 6.67(8) & 0.78(6) & 1.21(87) \\
33$^b$ & 40020-01-01-020$^b$ & ..... & 0.56(2) & 78(9)  & -2.1(2)  & 0.008(2)& 0.61(2) & 1.01(1) & 2.79(3) & 0.43(6) & 2.59(2) & 6.71(2) & 0.81(7) & 0.98(87) \\
33$^c$ & 40020-01-01-020$^c$ & ..... & 0.42(3) & 91(8)  & -2.0(3)  & 0.009(1)& 0.61(4) & 1.03(2) & 2.78(3) & 0.44(5) & 2.57(4) & 6.72(4) & 0.80(9) & 1.15(87) \\
33$^d$ & 40020-01-01-020$^d$ & ..... & 0.40(4) & 80(9)  & -2.3(2)  & 0.009(1)& 0.63(4) & 1.02(1) & 2.78(3) & 0.41(2) & 2.56(3) & 6.69(9) & 0.75(7) & 1.20(87) \\
33$^e$ & 40020-01-01-020$^e$ & ..... & 0.40(3) & 94(9)  & -2.1(3)  & 0.009(1)& 0.59(3) & 1.03(2) & 2.79(3) & 0.40(5) & 2.62(4) & 6.77(7) & 0.81(8) & 1.19(87) \\

34$^a$ & 40020-01-01-030$^a$ & 51188.260 & 0.44(4) & 100(10)& -2.1(2)  & 0.010(1)& 0.7(1)  & 1.01(1) & 2.55(1) & 0.40(2) & 2.61(2) & 6.97(7) & 0.50(6) & 1.18(87) \\
34$^b$ & 40020-01-01-030$^b$ & ..... & 0.43(5) & 100(10)& -2.0(3)  & 0.009(1)& 0.7(1)  & 1.01(2) & 2.56(1) & 0.42(5) & 2.60(4) & 6.94(6) & 0.49(8) & 1.20(87) \\
34$^c$ & 40020-01-01-030$^c$ & ..... & 0.44(4) & 110(10)& -2.0(2)  & 0.008(2)& 0.7(2)  & 1.03(1) & 2.54(2) & 0.41(4) & 2.63(3) & 6.96(9) & 0.50(7) & 1.16(87) \\
34$^d$ & 40020-01-01-030$^d$ & ..... & 0.43(2) & 110(10)& -2.1(3)  & 0.008(1)& 0.6(2)  & 1.02(1) & 2.56(1) & 0.40(5) & 2.64(2) & 6.78(8) & 0.49(7) & 1.18(87) \\
34$^e$ & 40020-01-01-030$^e$ & ..... & 0.43(4) & 110(10)& -2.2(3)  & 0.009(3)& 0.7(1)  & 1.01(2) & 2.55(2) & 0.40(6) & 2.65(3) & 6.62(9) & 0.50(8) & 1.17(87) \\
34$^f$ & 40020-01-01-030$^f$ & ..... & 0.43(4) & 100(10)& -2.1(2)  & 0.009(2)& 0.6(1)  & 1.01(1) & 2.56(3) & 0.39(5) & 2.65(2) & 6.56(9) & 0.47(6) & 1.18(87) \\

35$^a$ & 40020-01-01-04$^a$  & 51188.804 & 0.39(3) & 110(12)& -1.7(2)  & 0.04(1) & 0.60(3) & 0.97(2) & 2.75(1) & 0.33(1) & 2.69(2) & 6.49(9) & 0.92(5) & 1.08(87) \\
35$^b$ & 40020-01-01-04$^b$  & ..... & 0.38(2) & 110(15)& -1.7(1)  & 0.03(3) & 0.61(2) & 0.98(3) & 2.76(2) & 0.35(2) & 2.67(3) & 6.47(8) & 0.94(6) & 1.09(87) \\
35$^c$ & 40020-01-01-04$^c$  & ..... & 0.38(3) & 120(12)& -1.8(2)  & 0.02(1) & 0.60(1) & 0.96(1) & 2.75(1) & 0.34(1) & 2.66(4) & 6.48(5) & 0.93(7) & 1.09(87) \\
35$^d$ & 40020-01-01-04$^d$  & ..... & 0.37(4) & 120(15)& -1.9(3)  & 0.03(2) & 0.63(2) & 0.98(3) & 2.73(1) & 0.35(2) & 2.68(3) & 6.47(8) & 0.92(6) & 1.08(87) \\

36$^a$ & 40020-01-01-05$^a$  & 51189.152 & 0.41(2) & 120(10)& -2.1(2)  & 0.04(1) & 0.63(2) & 1.01(2) & 2.77(2) & 0.32(3) & 2.69(4) & 6.46(7) & 0.93(7) & 1.15(87) \\
36$^b$ & 40020-01-01-05$^b$  & ..... & 0.38(4) & 100(15)& -2.0(3)  & 0.03(2) & 0.62(1) & 1.02(1) & 2.79(3) & 0.30(3) & 2.66(3) & 6.47(8) & 0.90(5) & 1.25(87) \\
36$^c$ & 40020-01-01-05$^c$  & ..... & 0.39(4) & 110(10)& -1.9(4)  & 0.03(1) & 0.63(2) & 1.00(1) & 2.78(1) & 0.31(2) & 2.66(4) & 6.46(7) & 0.92(8) & 1.28(87) \\

37$^a$ & 40020-01-01-060$^a$ & 51191.785 & 0.98(4) & 10(3)  & -0.09(3) & 0.93(1) & 1.17(7) & 1.01(1) & 3.04(6) & 1.89(2) & 2.6(1)  & 6.47(7) & 0.42(5) & 1.02(87) \\
37$^b$ & 40020-01-01-060$^b$ & ..... & 0.38(6) & 90(9)  & -2.1(2)  & 1.02(1) & 0.52(4) & 1.03(2) & 2.68(1) & 0.57(3) & 2.68(4) & 6.59(3) & 0.89(7) & 1.23(87) \\
37$^c$ & 40020-01-01-060$^c$ & ..... & 0.37(4) & 110(10)& -2.1(2)  & 0.06(1) & 0.54(3) & 1.01(3) & 2.65(2) & 0.56(2) & 1.64(3) & 6.49(4) & 0.68(4) & 1.26(87) \\
37$^d$ & 40020-01-01-060$^d$ & ..... & 0.98(3) & 23(5)  & -2.1(2)  & 0.02(1) & 1.14(2) & 1.03(2) & 2.60(1) & 1.45(4) & 1.57(2) & 6.52(3) & 0.83(9) & 1.16(87) \\
37$^e$ & 40020-01-01-060$^e$ & ..... & 0.99(6) & 15(5)  & -0.06(2) & 0.04(1) & 1.15(3) & 1.02(2) & 2.08(5) & 1.74(2) & 0.7(1)  & 6.45(5) & 0.76(7) & 1.19(87) \\

38 & 40020-01-01-07  & 51192.852 & 0.41(5) & 120(10)& -2.2(4)  & 0.06(4) & 0.59(5) & 1.01(2) & 2.94(1) & 0.60(3) & 2.69(4) & 6.5(2)  & 0.43(6) & 1.08(87) \\
39 & 40020-01-03-00  & 51193.787 & 0.80(3) & 80(10) & -2.0(2)  & 0.05(3) & 0.51(3) & 1.02(3) & 3.01(2) & 0.82(9) & 2.68(3) & 6.9(1)  & 0.32(8) & 1.12(87) \\
40 & 40020-01-03-01  & 51194.860 & 0.82(2) & 69(7)  & -2.1(3)  & 0.06(4) & 0.45(4) & 0.99(1) & 2.95(1) & 0.69(5) & 2.68(3) & 6.5(1)  & 0.39(7) & 0.93(87) \\
41 & 40706-02-01-00  & 51339.384 & 0.98(4) & 2.0(3) & -0.08(3) & 1.4(1)  & 1.17(8) & 1.01(1) & 3.04(9) & 2.00$^{\dagger\dagger}$  & 0.6(1)  & 6.27(9) & 0.42(5) & 1.00(87) \\
42 & 40706-02-03-00  & 51339.642 & 1.02(1) & 3(1)   & -0.14(9) & 1.4(1)  & 1.22(9) & 0.98(2) & 3.01(8) & 2.00$^{\dagger\dagger}$  & 0.8(1)  & 6.23(8) & 0.42(6) & 1.06(87) \\
43 & 40706-02-06-00  & 51339.976 & 1.05(2) & 7.6(5) & -1.83(6) & 0.96(9) & 0.84(7) & 1.03(1) & 2.59(1) & 2.00$^{\dagger\dagger}$  & 1.14(2) & 6.39(5) & 0.65(4) & 1.29(87) \\
44 & 40706-01-01-000 & 51340.092 & 1.02(1) & 2.3(2) &  0.26(9) & 1.58(5) & 1.6(1)  & 1.01(1) & 3.4(1)  & 2.00$^{\dagger\dagger}$  & 0.3(1)  & 6.3(1)  & 0.42(5) & 1.33(87) \\
45 & 40706-02-08-00  & 51340.908 & 0.99(3) & 8.2(7) & -1.83(7) & 1.15(8) & 1.09(4) & 0.99(2) & 2.88(1) & 2.00$^{\dagger\dagger}$  & 1.38(1) & 6.18(8) & 0.46(5) & 0.76(87) \\
46 & 40706-02-09-00  & 51340.939 & 1.00(1) & 6.4(6) & -1.6(1)  & 1.11(9) & 1.05(3) & 1.02(3) & 2.85(2) & 2.00$^{\dagger\dagger}$  & 1.35(2) & 6.22(7) & 0.44(6) & 1.08(87) \\
47 & 40706-02-10-00  & 51340.974 & 1.03(2) & 6.5(4) & -1.58(5) & 1.16(8) & 1.08(4) & 1.03(2) & 2.84(2) & 2.00$^{\dagger\dagger}$  & 1.36(1) & 6.18(9) & 0.45(5) & 1.12(87) \\
48 & 40706-02-12-00  & 51341.155 & 1.07(4) & 5.7(3) & -0.45(9) & 1.17(9) & 1.07(3) & 1.01(1) & 2.82(3) & 2.00$^{\dagger\dagger}$  & 1.34(2) & 6.19(7) & 0.46(6) & 0.92(87) \\
49 & 40706-02-13-00  & 51341.222 & 0.99(2) & 5.2(2) & -0.4(2)  & 1.09(8) & 1.02(5) & 0.98(2) & 2.78(4) & 2.00$^{\dagger\dagger}$  & 1.32(3) & 6.23(8) & 0.44(7) & 0.93(87) \\
50 & 40706-02-14-00  & 51341.288 & 0.97(3) & 4.53(9)& -0.07(9) & 1.2(1)  & 1.10(6) & 0.99(1) & 2.79(7) & 2.00$^{\dagger\dagger}$  & 1.31(9) & 6.18(8) & 0.46(6) & 0.78(87) \\
51 & 40706-01-02-00  & 51341.355 & 1.00(2) & 2.3(2) &  0.2(1)  & 1.62(7) & 1.5(1)  & 1.01(1) & 3.4(1)  & 2.00$^{\dagger\dagger}$  & 0.4(1)  & 6.32(7) & 0.43(5) & 1.34(87) \\
52 & 40706-02-16-00  & 51341.576 & 1.03(1) & 3.3(2) & -0.27(8) & 1.28(9) & 1.02(8) & 0.98(3) & 2.17(9) & 2.00$^{\dagger\dagger}$  & 1.0(1)  & 6.32(6) & 0.54(5) & 1.04(87) \\
53 & 40706-02-17-00  & 51341.699 & 1.00(4) & 3.7(3) & -0.58(9) & 0.97(8) & 0.82(9) & 1.03(4) & 2.36(7) & 2.00$^{\dagger\dagger}$  & 0.98(9) & 6.32(7) & 0.49(4) & 1.03(87) \\
54 & 40706-02-18-00  & 51341.905 & 0.93(3) & 5.9(8) & -1.50(8) & 0.94(9) & 0.90(6) & 1.02(1) & 2.71(2) & 2.00$^{\dagger\dagger}$  & 1.26(2) & 6.32(7) & 0.45(5) & 0.94(87) \\
55 & 40706-02-19-00  & 51341.972 & 1.02(4) & 6.6(5) & -1.5(3)  & 0.58(7) & 0.67(8) & 0.99(2) & 2.67(1) & 2.00$^{\dagger\dagger}$  & 1.22(9) & 6.38(8) & 0.43(5) & 1.23(87) \\
56 & 40706-02-21-00  & 51342.155 & 1.07(2) & 7.8(6) & -1.6(1)  & 0.5(1)  & 0.80(6) & 1.01(1) & 2.65(3) & 2.00$^{\dagger\dagger}$  & 1.76(3) & 6.69(7) & 0.54(4) & 1.18(87) \\
57 & 40706-02-23-00  & 51342.287 & 0.99(2) & 2.2(1) &  0.03(1) & 1.4(2)  & 1.2(1)  & 0.98(2) & 2.89(9) & 2.00$^{\dagger\dagger}$  & 0.68(9) & 6.32(6) & 0.50(7) & 0.83(87) \\
58 & 40706-01-03-00  & 51342.354 & 1.01(3) & 2.0(1) &  0.26(9) & 1.49(8) & 1.25(9) & 1.01(1) & 3.03(8) & 2.00$^{\dagger\dagger}$  & 0.6(1)  & 6.39(7) & 0.50(5) & 1.18(87) \\
59 & 70015-01-01-00  & 52413.913 & 1.01(4) & 4.9(3) & -1.9(2)  & 2.0(1)  & 0.59(8) & 1.02(3) & 2.78(2) & 0.63(9) & 2.23(5) & 6.41(9) & 0.4(1)  & 0.86(87) \\
60 & 70015-01-01-01  & 52413.982 & 1.08(3) & 6.5(4) & -1.9(1)  & 2.01(9) & 0.68(7) & 0.99(1) & 2.68(2) & 0.6(1)3 & 2.09(4) & 6.3(1)  & 0.7(1)  & 1.17(87) \\
     \enddata
    \label{tab:fit_table_rxte}
$^\dagger$ The spectral model is  $wabs*(Comptb1 + Comptb2 + Gaussian)$;
$^{\dagger\dagger}$ when parameter $\log(A_2)\gg1$, this parameter is fixed at 2.0 (see comments in the text), 
$^{\dagger\dagger\dagger}$ normalization parameters of 
$COMPTB$ components are in units of 
$L_{39}/d^2_{10}$, where $L_{39}$ is the source luminosity in units of 10$^{39}$ erg/s, 
$d_{10}$ is the distance to the source in units of 10 kpc 
and $Gaussian$ component is in units of $10^{-2}\times total~~photons$ $cm^{-2}s^{-1}$ in line, 
$\sigma_{line}$ of Gaussian component is fixed to a value 0.8 keV (see comments in the text),
$N_H$ was fixed at value of 3$\times 10^{21}$ cm$^{-2}$ (Christian \& Swank, 1997). 
 Some data sets, although having the same proposal number, contain 
observation separated by time intervals. In these cases we collected all the 
observations close 
in just one CCD interval (see Fig.~1), and distinguish the relative CCDs by a superscripts (e.g., ``a'', ``b'' etc), 
following the proposal number
\end{deluxetable}

\vspace{2.in}
\newpage
~~~~~~~~
\begin{deluxetable}{llccccccc}
\tablewidth{0in}
\tabletypesize{\scriptsize}
    \tablecaption{Comparisons of the best-fit parameters  of {\it Z}-sources Sco~X-1 and GX~340+0$^1$ and {\it atoll} 
sources GX~3+1$^2$, 4U~1728-34$^3$, 4U1820-30$^4$ and ``$atoll$+Z'' source XTE~J1701-462$^5$}
    \renewcommand{\arraystretch}{1.2}
 \tablehead
{Source & Alternative & Class$^4$& Distance, & Presence of & $kT_e$,  & $ N_{comptb}$ &  $kT_{s}$  & $f$ \\
  name  & name        &          & kpc       & kHz QPO     & keV         &  $L_{39}^{soft}/{D^2_{10}}$        & keV  &  }
 \startdata
4U~1617-15  & Sco~X-1  & Z, Sp, B     & 2.8$^{7}$ &    +$^{8}$        & 3-180 &  0.3-3.4 & 0.4-1.8 & 0.08-1   \\
4U~1642-45  & GX 340+0 & Z, Sp, B     & 10.5$^{9}$ &    +$^{12}$        & 3-21   &  0.08-0.2  & 1.1-1.5 & 0.01-0.5   \\
4U~1744-26  & GX 3+1   & Atoll, Sp, B & 4.5$^{10}$ &     none$^{13}$      & 2.3-4.5 &  0.04-0.15 & 1.16-1.7 & 0.2-0.9   \\
4U~1728-34  & GX~354-0 & Atoll, Su, D & 4.2-6.4$^{11}$ & +$^{14}$&  2.5-15& 0.02-0.09 &  1.3 & 0.5-1\\     
4U~1820-30  &   ...    & Atoll, Su, - & 5.8-8$^{15}$& +$^{16}$&  2.9-21& 0.02-0.14 &  1.1-1.7 & 0.2-1\\     
XTE~J1701-462&  ...    & Atoll+Z, Su, - & 8.8$^{17}$& +$^{18}$&  ...   & ...       &  1-2.7 & ...\\     
      \enddata
    \label{tab:fit_table_comb}
References:
(1) STF13;
(2) ST12; 
(3) ST11;  
(4) TSF13; 
(5) LRH109;
(6) Classification of the system in the various schemes (see text): Sp = supercritical, Su = subcritical, 
B = bulge, D = disk; 
(7) Bradshaw et al. (1999); 
(8) Zhang et al. (2006); 
(9) Fender \& Henry  (2000),
Christian \& Swank (1997); 
(10) \citet{kk00}  
(11) \citet{par78};  
(12) \citet{Jonker98};  
(13) \citet{stroh98}; 
(14) \citet{to99} 
(15) \citet{ST04} 
(16) \citet{Smale97}
(17) \citet{Lin07a}, \citet{Lin09}
(18) \citet{Sana10}
\end{deluxetable}

\newpage

%
%

\newpage

\newpage


\newpage


\newpage


\newpage

%
%

\newpage 

%
%

\newpage

%
%

\newpage

%
%

\newpage

%
%

\newpage

%
%

\newpage

%
%

\newpage

%
%

\newpage 

%
%

\end{document}